\documentclass[usegraphicx,usenatbib]{mn2e}

\title[A Test for the Disruption of Magnetic Braking]{How Many CVs are Crossing the Period Gap? A Test for the Disruption of Magnetic Braking}

\author[P. J. Davis et al.]{P.~J.~Davis,$^1$ U.~Kolb,$^1$ B.~Willems,$^2$
 B.~T.~G\"{a}nsicke,$^3$ \\ $^1$The Open University, Department of
 Physics and Astronomy, Walton Hall, Milton Keynes MK7 6AA \\
 $^2$Northwestern University, Department of Physics and Astronomy,
 2131 Tech Drive, Evanston, IL 60208, USA \\ $^3$The University of
 Warwick, Department of Physics and Astronomy, Coventry CV4 7AL}
\begin{document}

\maketitle

\begin{abstract}
We apply population synthesis techniques to calculate the present day
number of two types of white dwarf-main sequence star (WDMS) binaries
within the cataclysmic variable 2 to 3 hour period gap. The first are
post-common envelope binaries with secondary stars that have masses
$0.17\le{M_{2}/\rmn{M_{\odot}}}\le{0.36}$ (gPCEBs), such that they
will commence mass transfer within the period gap. The second type are
systems that were CVs at some point in their past, but detached once
they evolved down in orbital period to $\approx$ 3 h as a consequence
of disrupted magnetic braking, and are crossing the period gap via
gravitational radiation (dCVs). Full population synthesis calculations
are performed where we assume either constant, global values of the
common envelope ejection efficiency, $\alpha_{\rmn{CE}}$, or consider
$\alpha_{\rmn{CE}}$ as a function of secondary mass. Several forms of
magnetic braking are also considered. We predict an excess of dCVs
over gPCEBs within the period gap of $\sim$4 to $\sim$13 assuming
$\alpha_{\rmn{CE}}=0.1-0.6$, and an initial mass ratio distribution of
the form $n(q)=1$. This excess is revealed as a prominent peak at the
location of the period gap in the orbital period distribution of the
combined gPCEB and dCV population. We suggest that if such a feature
is observed in the orbital period distribution of an observed sample
of short orbital period WDMS binaries, this would strongly corroborate
the disruption of magnetic braking.
\end{abstract}


\section{Introduction}
The standard paradigm of cataclysmic variable (CV) evolution assumes
that orbital angular momentum (AM) is lost via both gravitational
radiation and magnetic braking at orbital periods
$P_{\rmn{orb}}/\rmn{h}>{3}$, and purely by gravitational radiation for
$P_{\rmn{orb}}/\rmn{h}\la{3}$ (e.g. Warner 1995; King 1988). However,
the AM loss rate associated with magnetic braking is ill-constrained,
and it is unclear if magnetic braking ceases to be a sink of AM for
$P_{\rmn{orb}}/\rmn{h}\la{3}$ \citep{kolb02}.

Extrapolating the work of \citet{skumanich72} who investigated the
spin down rates of G-type stars by a magnetically coupled stellar
wind, \citet{vz81} obtained the magnetic braking AM loss rate of the
form $\dot{J}_{\rmn{MB}}\propto{\Omega}^{3}$, where $\Omega$ is the
spin frequency of the donor star.

As a consequence of the AM losses and the mass loss they drive, the
donor star is unable to maintain thermal equilibrium. This causes
donors with masses $\la{0.8}$ M$_{\odot}$ to exceed their equilibrium
radius \citep*{srk96}. Mass loss will eventually cause the donor star
to become fully convective at a mass $M_{2}= M_{2,\rmn{conv}}$ and
when $P_{\rmn{orb}}/\rmn{h}\approx{3}$. At this point it is suggested
that either the magnetic activity of the donor star discontinuously
decreases, or the topology of the magnetic field changes
\citep{ts89}. This is based on the result that the interface between
the radiative core and the convective envelope play a key role in the
generation of the magnetic field \citep{mt89}.

When magnetic braking ceases at $M_{2}={M_{2,\rmn{conv}}}$ the donor
shrinks within its Roche lobe, mass transfer is shut off and the
secondary re-attains thermal equilibrium. The system is no longer
observed as a CV. Subsequent orbital evolution of the CV is driven by
gravitational radiation only. Mass transfer resumes again when the
secondary star makes contact with its Roche lobe once more at
$P_{\rmn{orb}}/\rmn{h}\approx{2}$.

\begin{figure}
  \begin{center}
%  \vspace{174pt}
  \includegraphics[scale=0.4]{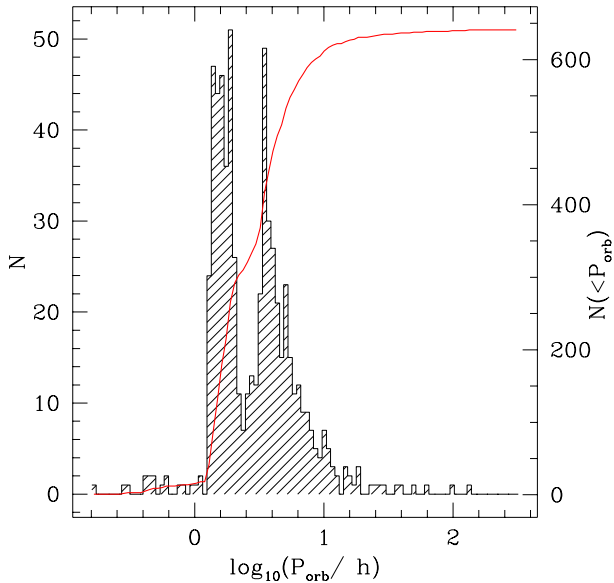}
  \end{center}
  \caption{Orbital period distribution of CVs from Ritter \& Kolb
  (2003), catalogue edition 7.7 (2006). The axis on the left indicates
  the number of systems per bin. The solid, red curve is the
  cumulative distribution with numbers, $N<P_{\rmn{orb}}$, indicated
  on the right axis.}
  \label{Pdist}
\end{figure}

This disrupted magnetic model was conceived to explain the lack of
observed CVs in the range $2\la{P_{\rmn{orb}}/\rmn{h}}\la{3}$
(Rappaport, Verbunt \& Joss 1983; Spruit \& Ritter 1983), as shown in
Figure \ref{Pdist}.

However, despite the success of this model in reproducing the correct
width and position of the period gap (e.g. Kolb, King \& Ritter 1998),
there is no independent observational support for disrupted magnetic
braking. Observations of young, open clusters (Sills, Pinsonneault \&
Terndrup 2000) show that the spin down rate of rapidly rotating single
stars is much less than predicted by the \citet{vz81} relation. Also,
no discontinuous decline in magnetic activity for the fully convective
stars is detected (Jones, Fischer \& Stauffer 1996). It is likely that
for the high rotational speeds of $\ga{100}$ km s$^{-1}$ of the CV
donor stars, the AM loss rate, $\dot{J}_{\rmn{MB}}$, due to magnetic
braking will saturate (for a review see e.g. Collier-Cameron
2002). \citet*{spt00} suggest the form
$\dot{J}_{\rmn{MB}}\propto{\Omega}$ in the saturated regime. Within
this scheme the AM loss rate in CVs would be a factor of
100 less than that for the standard magnetic braking model, and the
donor star would not be sufficiently driven out of thermal equilibrium
and no period gap would arise \citep*{aps03}.

Such theoretical and observational problems with the standard model of
CV evolution aside, it is the only one thus far that predicts a period
gap of the correct width with edges at the correct orbital periods
\footnote{Willems et al. (2005) and Willems et al. (2007) studied CV
populations evolving under the influence of circumbinary disks and
found a period gap feature, but had to appeal to additional mechanisms
in the pre-CV evolution to account for the lower edge of the gap.}. Two
recent pieces of evidence in support for the disrupted magnetic
braking model lies in the mass-radius relation of donor stars in CVs
(Smith \& Dhillon 1998; Patterson et al. 2005; Knigge 2006) and in the
effective temperatures of the accretion-heated white dwarfs in CVs
(Townsley \& Bildsten 2003). In the former, the donor mass-radius
relation for $P_{\rmn{orb}}/\rmn{h}\ga{3}$ lies systematically above
that for isolated main sequence stars. This is consistent with the
donor star being driven out of thermal equilibrium. Furthermore, a
discontinuous decrease in the donor radii between
$2\la{P_{\rmn{orb}}}/\rmn{h}\la{3}$ has been detected. In the latter
piece of evidence, the effective temperatures of white dwarfs are
higher above the period gap than below, and are overall consistent
with the expected accretion heating for mass transfer rates predicted
by the disrupted magnetic braking model.

Politano \& Weiler (2006) calculated the secondary star mass
distribution of present day PCEBs for a range of AM loss
prescriptions, including disrupted magnetic braking.  They found that
the number of PCEBs per secondary mass interval decreases by 38 per
cent once magnetic braking begins to drive their evolution for
secondaries with radiative cores. They suggest that this decrease in
the number of PCEBs with secondaries between $\sim$ 0.25 and 0.5
M$_{\odot}$ provides an observable test for the disruption of magnetic
braking.

In this paper we suggest an alternative, even more direct, test for
the disruption of magnetic braking by investigating a more explicit
consequence: the existence of a population of WDMS binaries that were
CVs in the past but became detached at
$P_{\rmn{orb}}/\rmn{h}\approx{3}$ and are currently crossing the
period gap. We call these dCVs.

Using population synthesis techniques we calculate the present day
population of these dCVs, as well as WDMS binaries that have
emerged from a common envelope (CE) phase. We call these systems
post-CE binaries (PCEBs). In particular, we only consider PCEBs which
will begin mass transfer within the period gap. For our stellar models
and choice of gap boundaries, such systems have secondary masses in
the range $0.17\le{M_{2}/\rmn{M}_{\odot}}\le{0.36}$. This sub-set of
PCEBs we henceforth designate as gPCEB systems. Note that gPCEBs have
$P_{\rmn{orb}}/\rmn{h}\ge{2}$ but can also have
$P_{\rmn{orb}}/\rmn{h}>3$.

% Using population synthesis we predict the present day population of
%two types of detached white dwarf-main sequence (WDMS) star
%binaries. The first type are detached WDMS systems that have emerged
%from a common envelope (CE) phase. We call these post-CE binaries
%(PCEBs). In particular, we only consider PCEBs which will begin mass
%transfer within the period gap. For our stellar models and choice of
%gap boundaries, such systems have secondary masses in the range
%$0.17\le{M_{2}/\rmn{M}_{\odot}}\le{0.36}$. This sub-set of PCEBs we
%henceforth designate as gPCEB systems. Note that gPCEBs have
%$P_{\rmn{orb}}/\rmn{h}\ge{2}$ but can also have
%$P_{\rmn{orb}}/\rmn{h}>3$. The second variety are WDMS star binaries
%that were CVs at some point in their past but became detached at
%$P_{\rmn{orb}}/\rmn{h}\approx{3}$ and are currently crossing the
%period gap. We loosely call these detached CVs (dCVs).}

As we will show below orbital the period distribution of the combined
population of dCVs and gPCEBs reveals a prominent peak within the
period gap due to an excess of dCVs over gPCEBs here. We calculate the
ratio dCV:gPCEB within the period gap, which indicates the height of
this peak, for a range of models concerning the values of the CE
ejection efficiency, $\alpha_{\rmn{CE}}$, the initial distribution of
secondary mass, and the AM loss rate due to magnetic braking.  The
detection of such a feature in the orbital period distribution of an
observed sample of WDMS binaries with orbital periods of a few hours
would provide a simple test for the disruption of magnetic braking.

The Sloan Digital Sky Survey (SDSS) is currently creating the
opportunity for large-scale observational studies of WDMS binaries,
having identified already more than 1000 systems (Silvestri et
al. 2007; Schreiber, Nebot Gomez-Moran \& Schwope 2007). Attempts to
establish the subsample of PCEBs among those systems and measuring
their orbital periods as well as their stellar properties is underway
(Rebassa-Mansergas et al. 2007, Schreiber et al. 2008). We are
optimistic that a quantitative test of the calculations carried out in
this paper, and hence of the validity of the disrupted magnetic
braking hypothesis, will be feasible in the near future.

The structure of the paper is as follows. In Section 2 we describe the
computational method, in Section 3 we present our simulation results,
which are discussed in Section 4. We conclude our investigation in
Section 5.

\section{Computational Method}

The computational method we apply consists of two main steps; the
first is to calculate the zero-age populations of the dCVs (i.e. CVs
that have just detached at the upper edge of the period gap) and PCEBs
(i.e. WDMS binaries that have just emerged from a CE phase) using
the population synthesis code BiSEPS (Willems \& Kolb 2002; Willems \&
Kolb 2004). These populations are unweighted i.e. no formation
probabilities are yet calculated for the systems. The code employs the
single star evolution (SSE) formulae described in \citet*{hpt00} and a
binary system evolution scheme based on that described in
\citet*{hpt02}. The second step uses a different code introduced in
Willems et al. (2005), which calculates the present day populations of
these systems using a library of evolutionary tracks, here produced by
BiSEPS itself. The procedure is described in more detail in the
following sections.

\subsection{Initial Binary Population}

BiSEPS evolves a large number of binary systems, which initially
consist of two zero-age main sequence stellar components. The stars
are assumed to have a population I chemical composition and the orbits
are circular at all times. The initial primary and secondary masses
are in the range 0.1 to 20 M$_{\odot}$, while the initial orbital
periods range from 0.1 to 100 000 d. There is one representative
binary configuration per grid cell within a three-dimensional grid
consisting of 60 logarithmically spaced points in primary and
secondary mass and 300 logarithmically spaced points in orbital
period. There are $\sim{5.4\times{10}^{6}}$ binaries that are evolved
for a maximum evolution time of 10 Gyr. For symmetry reasons only
systems with $M_{1}>M_{2}$ are evolved.

We calculate the four dimensional probability density functions
$\chi(t,M_{\rmn{WD}},M_{2},P_{\rmn{orb}})$ of all zero-age dCVs and
$\phi(t,M_{\rmn{WD}},M_{2},P_{\rmn{orb}})$ for all zero-age PCEBs
using the initial distribution functions quoted below. Here $t$ is the
time since the birth of the Galaxy, $M_{\rmn{WD}}$ is the mass of the
white dwarf, $M_{2}$ is the mass of the secondary star and
$P_{\rmn{orb}}$ is the orbital period. The number of zero-age main
sequence primaries, $\rmn{d}N$, that form with masses in the range
$\rmn{d}M_{1}$ is given by $\rmn{d}N=f(M_{1})\rmn{d}M_{1}$, where the
probability density function, $f(M_{1})$, is given by

\begin{equation}f(M_{1}) = \left\{
\begin{array}{l l}
  0 & \quad \mbox{$M_{1}/\rmn{M_{\odot}}<0.1,$}\\ 0.29056M_{1}^{-1.3}
  & \quad \mbox{$0.1\leq{M_{1}/\rmn{M_{\odot}}}<0.5,$}\\
  0.15571M_{1}^{-2.2} & \quad
  \mbox{$0.5\leq{M_{1}/\rmn{M_{\odot}}}<1.0,$}\\ 0.15571M_{1}^{-2.7} &
  \quad \mbox{$1.0\leq{M_{1}/\rmn{M_{\odot}}},$}\\ \end{array}
  \right.
\label{M1dist}\end{equation}
(Kroupa, Tout \& Gilmore 1993). The probability that a zero-age main
sequence secondary star forms with a mass $M_{2}$ can be determined
from the initial mass ratio distribution (IMRD) given by

\begin{equation}n(q) = \left\{
\begin{array}{l l}
  \mu{q^{\nu}} & \quad \mbox{$0<q\leq{1},$}\\ 0 & \quad
  \mbox{$q>1,$}\\ \end{array} \right. \label{qdist}\end{equation}
  where $q=M_{2}/M_{1}$, $\nu$ is a constant, and $\mu$ is a
  normalisation factor such that $\int{n(q)}\,\rmn{d}q=1$ ($\mu=\nu+1$
  for $\nu>-1$). We also consider the case where the secondary mass is
  determined from an IMF according to equation (\ref{M1dist}) where
  $M_{1}$ is replaced with $M_{2}$. We use $n(q)=1$, i.e. $\nu=0$, as
  our reference model. Finally, the probability that a binary forms
  with an initial orbital separation $a$ is determined by (Iben \&
  Tutukov 1984; Hurley et al. 2002)
\begin{equation}h(a) = \left\{
\begin{array}{l l}
  0 & \quad \mbox{$a/\rmn{R_{\odot}}<3\:\:\rmn{or}\:\:a/\rmn{R_{\odot}}>10^{6},$}\\
  0.078636a^{-1} & \quad \mbox{$3\leq{a/\rmn{R_{\odot}}}\leq{10^{6}}.$}\\ \end{array} \right.
\label{adist}\
\end{equation}

We further assume that all stars in the Galaxy are formed in binaries
and that the Galaxy has an age of 10 Gyr. The star formation rate,
$S$, is calculated by assuming that one binary per year is formed with
$M_{1}>0.8\:\rmn{M_{\odot}}$. Combining this with the Galactic volume
of $5\times{10}^{11}$ pc$^{3}$ gives an average local birthrate of
white dwarfs of $2\times{10}^{-12}$ pc$^{-3}$ yr$^{-1}$, consistent
with observations (Weidemann 1990). The lower limit of 0.8 M$_{\odot}$
is the smallest mass the primary star can have if it is to evolve into
a white dwarf within the lifetime of the Galaxy. We therefore have
\begin{equation}
S\int^\infty_{0.8}{f(M_{1})\,\rmn{d}M_{1}}=1\:\rmn{yr^{-1}}.
\label{SFR}
\end{equation}
From (\ref{M1dist}) and (\ref{SFR}) we obtain $S=7.6$ yr$^{-1}$. We
assume that this is constant throughout the lifetime of the Galaxy.

We then employ a population calculator to convert $\chi{(t,M_{\rmn
{WD}},M_{2},P_{\rmn{orb}})}$ and $\phi{(t,M_{\rmn{WD}},M_{2},P_{\rmn
{orb}})}$ into distributions over $M_{\rmn{WD}}$, $M_{2}$ and $P_{\rmn
{orb}}$ describing the present day population of dCVs and gPCEBs. The
secular evolution of each system in the zero-age populations is
therefore followed using BiSEPS, and its contribution to the
population, determined by the above distribution function and star
formation rate, is accounted for at each stage of its evolution.

%We then employ a population calculator that is in extension to BiSEPS,
%which utilises the evolutionary tracks in order to convert
%$\chi{(t,M_{\rmn{WD}},M_{2},P_{\rmn{orb}})}$ and
%$\phi{(t,M_{\rmn{WD}},M_{2},P_{\rmn{orb}})}$ into distributions over
%$M_{\rmn{WD}}$, $M_{2}$ and $P_{\rmn{orb}}$ describing the present day
%population of dCVs and gPCEBs.

\subsection{Binary Evolution}

The CE phase \citep{paczynski76} is modelled by equating the binding
energy of the primary's envelope to the change in the total orbital
energy of the binary system. Quantitatively this is
\begin{equation}
\frac{GM_{\rmn{e}}(M_{\rmn{e}}+M_{\rmn{c}})}{\lambda{R_{1,\rmn{L}}}}=\alpha_{\rmn{CE}}\left[\frac{GM_{\rmn{c}}M_{2}}{2a_{\rmn{f}}}-\frac{G(M_{\rmn{c}}+M_{\rmn{e}})M_{2}}{2a_{\rmn{i}}}\right]
\label{ce},\end{equation} where $G$ is the gravitational constant,
$M_{\rmn{c}}$ and $M_{\rmn{e}}$ are the masses of the primary's core
(the proto-white dwarf) and envelope, $R_{1,\rmn{L}}$ is the radius of
the primary immediately before the onset of the CE phase, and $M_{2}$
is the mass of the secondary. The orbital separation of the stellar
components immediately before and after the CE phase are $a_{\rmn{i}}$
and $a_{\rmn{f}}$ respectively. The constant $\alpha_{\rmn{CE}}$ is
the fraction of the orbital energy used to unbind the envelope from
the core of the primary and $\lambda$ is the ratio between the
approximate form of the binding energy
$GM_{\rmn{e}}(M_{\rmn{e}}+M_{\rmn{c}})/R_{1,\rmn{L}}$ and the true
binding energy. Throughout our investigation we use $\lambda=0.5$,
which is consistent with the value used by Willems \& Kolb (2004). We
take $\alpha_{\rmn{CE}}=1.0$ as our reference model `A'.

Once the CE has been ejected from the system, then if the binary's
stellar components have avoided a collision, what remains is a PCEB
system. The subsequent evolution of the system is driven by orbital AM
losses; a combination of magnetic braking and gravitational radiation
if the secondary mass is larger than the mass of a fully convective,
isolated main sequence star, $M_{\rmn{MS,conv}}$, and purely by
gravitational radiation otherwise.

The system will become semi-detached once the orbital separation has
decreased sufficiently so that the secondary star is brought into
contact with its Roche lobe. If the mass ratio $q\la{1.3}$ then mass
transfer driven by AM losses will ensue. If, however, $q\ga{1.3}$ then
the secondary will undergo thermal timescale mass transfer
(TTMT). Depending on the initial mass ratio and evolutionary state,
some of these systems may reappear as AM-driven CVs. It is thought
that as many as 40 per cent \citep{kw05} of the present population of
CVs may have formed in this way. The detection of CNO abundance
anomalies supports this hypothesis, placing a lower limit of 10 per
cent on the fraction of post-TTMT systems among the observed
population of CVs (G\"ansicke et al. 2003). Determination of the
parameters at which the TTMT phase reverts to the standard AM-driven
phase would require a comprehensive treatment of the TTMT phase, and
would depend on the fate of the transferred material (Schenker
2001). We therefore only consider the contribution of purely AM driven
CVs in the present investigation.

\subsubsection{The Ejection Efficiency Parameter, $\alpha_{\rmn{CE}}$}

\begin{figure}
  \begin{center}
    \includegraphics[scale=0.45]{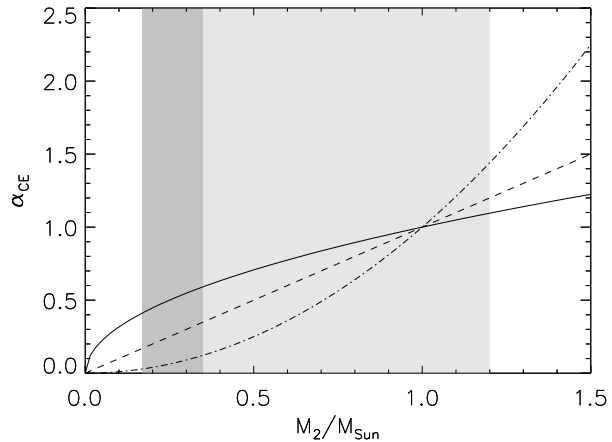}
  \end{center}
  \caption{Variation of $\alpha_{\rmn{CE}}$ with $M_{2}$ according to
    eqn. (\ref{alpha1}). Solid: $p=0.5$; dashed: $p=1.0$; dot-dashed:
    $p=2.0$. Dark-grey shading indicates range of secondary masses in
    gPCEB systems; light-grey range of secondary masses for dCV progenitors.}
  \label{alpha_m2}
\end{figure}

In our calculations, we consider the impact of changing the common
envelope ejection efficiency $\alpha_{\rmn{CE}}$. In addition to
constant, global values of $\alpha_{\rmn{CE}}$ we follow Politano \&
Weiler (2007) and consider the possibility that $\alpha_{\rmn{CE}}$
could be a function of the secondary mass. We adopt
\begin{equation}
\alpha_{\rmn{CE}}=\left(\frac{M_{2}}{\rmn{M_{\odot}}}\right)^p,
\label{alpha1}
\end{equation}
where $p$ is a free parameter. Figure \ref{alpha_m2} shows how
$\alpha_{\rmn{CE}}$ varies with $M_{2}$ for $p=0.5$, 1.0 and 2.0. The
dark-grey shading is the range of secondary masses in gPCEBs
considered here, while the light-grey shading is the range of
secondary masses found for dCV progenitors. Table \ref{models1}
summarises our population synthesis models for varying values of
$\alpha_{\rmn{CE}}$. In our model names, the first portion in lower
case denotes the form of magnetic braking used. In Table \ref{models1}
`h' denotes that the Hurley et al. (2002) form of magnetic braking is
used according to equation (\ref{hurley_mb}). The label `CE$x$'
denotes the value, $x$, of the constant, global value of
$\alpha_{\rmn{CE}}$. For example `CE01' is
$\alpha_{\rmn{CE}}=0.1$. Finally, the acronym `PWR$y$' denotes the use
of equation (6) with the value, $y$, of the parameter $p$, for example
`PWR05' is $p=0.5$.

\begin{table}
  \centering
%  \begin{minipage}{140mm}
    \caption{Model assumptions for $\alpha_{\rmn{CE}}$ and magnetic
    braking.}
%    \begin{tabular}{@{}llcl@{}}
    \begin{tabular}{@{}llc@{}}
    \hline
    Model   &   $\alpha_{\rmn{CE}}$          &  Magnetic braking law\\
    \hline
%    \hline
    hA      &   1.0                          &  eqn. (\ref{hurley_mb})\\
    hCE01   &   0.1                          &  eqn. (\ref{hurley_mb})\\
    hCE06   &   0.6                          &  eqn. (\ref{hurley_mb})\\
    hCE3    &   3.0                          &  eqn. (\ref{hurley_mb})\\
    hCE5    &   5.0                          &  eqn. (\ref{hurley_mb})\\
    hPWR05  &   eqn.(\ref{alpha1}), $p=0.5$  &  eqn. (\ref{hurley_mb})\\
    hPWR1   &   eqn.(\ref{alpha1}), $p=1.0$  &  eqn. (\ref{hurley_mb})\\
    hPWR2   &   eqn.(\ref{alpha1}), $p=2.0$  &  eqn. (\ref{hurley_mb})\\
\hline
\hline
    rvj2A   &   1.0                          &  eqn. (\ref{rvj_mb}), $\gamma=2$\\
    rvj4A   &   1.0                          &  eqn. (\ref{rvj_mb}), $\gamma=4$\\
\hline
\end{tabular}
\label{models1}
\end{table}

\subsubsection{Magnetic Braking}

We also investigate how various forms of magnetic braking affect our
results.~\citet{rvj83} give a parameterised form of magnetic braking
as
\begin{equation}
\dot{J}_{\rmn{rvj}}=-\eta_{\rmn{rvj}}\,3.8\times{10}^{-30}\,R_{\odot}^{4}\,{\rmn{s\:cm^{2}}}\:M_{2}\left(\frac{R_{2}}{R_{\odot}}\right)^{\gamma}\Omega^{3},
\label{rvj_mb}
\end{equation}
where $\Omega$ is the spin frequency of the secondary star and
$\gamma$ is a free parameter. If $\gamma=4$ then (\ref{rvj_mb})
reduces to the form derived by \citet{vz81}. The dimensionless
constant $\eta_{\rmn{rvj}}$ is determined so as to give the correct AM
loss rate at the upper edge of the period gap, which we determine in
Section 2.2.3. The prescription given by Hurley et al. (2002) is given
by
\begin{equation}
\dot{J}_{\rmn{h}}=-\eta_{\rmn{h}}\,5.83\times{10}^{-16}\:\rmn{M_{\odot}\:R_{\odot}^{-1}\:yr}\:\frac{M_{\rmn{env}}}{M_{2}}R_{2}^{3}\Omega^{3},
\label{hurley_mb}
\end{equation}
where $M_{\rmn{env}}$ is the mass of the secondary star's convective
envelope and $R_{2}$ is the donor's radius. Thus, magnetic braking
vanishes for a star with no convective envelope. As with equation
(\ref{rvj_mb}) the value of the constant $\eta_{\rmn{h}}$ is
determined to give the appropriate AM loss rate at the upper edge of
the period gap. We use equation (\ref{hurley_mb}) for our reference
model. Table 1 summarises our models where we use different forms of
magnetic braking. The acronym `rvj2' and `rvj4' denotes that we use
the Rappaport et al. (1983) form of magnetic braking with $\gamma=2$
and 4 respectively.

%\begin{table}
%  \centering
%%  \begin{minipage}{140mm}
%    \caption{Adopted magnetic braking prescriptions.}
%%    \begin{tabular}{@{}llcl@{}}
%    \begin{tabular}{@{}lll@{}}
%      \hline
%      Model  &  $\alpha_{\rmn{CE}}$  &   Magnetic braking law\\
%%      \hline
%      \hline
%      rvj2A  &  1.0                  &   eqn.(\ref{rvj_mb}), $\gamma=2$\\
%      rvj4A  &  1.0                  &   eqn.(\ref{rvj_mb}), $\gamma=4$\\
%      hA     &  1.0                  &   eqn.(\ref{hurley_mb})\\
%      \hline
%\end{tabular}
%\label{models2}
%\end{table}

\subsubsection{The Period Gap}

In this section we outline how we determine the position, and
therefore the width, of the period gap. We then ascertain the values
of $\eta_{\rmn{rvj}}$ and $\eta_{\rmn{h}}$, which will give the AM
loss rate at the upper edge of the period gap appropriate for its
determined width. Mass loss from the donor will cause it to exceed its
equilibrium radius (see Section 2.2.5) by a factor $f_{\rmn{conv}}$ at
the upper edge of the period gap. We calculate this value at the end
of this section.

We obtain the position of the gap edges by calculating a moving
average of the cumulative distribution curve shown in Figure
\ref{Pdist}. Linear fits were performed on three segments of the
curve; within and on either side of the period gap. Specifically, we
consider the curve in the range
$0.22\le{\rmn{log}_{10}}(P_{\rmn{orb}}/\:\rmn{h})\le{0.64}$. The edges
of the period gap are where these fits intersect. We find that the
upper and lower edges are located at 3.0 and 2.0 hours
respectively. We adopt these values throughout the rest of our
investigation.

The magnetic braking laws described in Section 2.2.2 are now
calibrated so as to give the correct AM loss rate, and therefore the
correct mass transfer rate, at the upper edge of the period gap. Kolb
(1996) determined that the mass transfer rate at the upper edge of the
period gap is $\approx{10}^{-9}$ M$_{\odot}$ yr$^{-1}$. For the
Rappaport et al. (1983) form of magnetic braking we find that
$\eta_{\rmn{rvj}}=0.28$ for $\gamma=2$. For $\gamma=4$ we find
$\eta_{\rmn{rvj}}=1.5$. For the Hurley et al. (2002) prescription,
$\eta_{\rmn{h}}=0.19$.

We now determine the value of $f_{\rmn{conv}}$. The Roche lobe radius
of the donor star, $R_{2,\rmn{L}}$, from Roche geometry
\begin{equation}
R_{2,\rmn{L}}\approx{0.44}\left(\frac{M_{2}}{M_{2}+M_{1}}\right)^{1/3}a,
\label{RL2}
\end{equation}
(Paczy\'{n}ski 1971) and Kepler's law gives
\begin{equation}
P_{\rmn{orb}}\propto\left(\frac{R_{2,\rmn{L}}^{3}}{M_{2}}\right)^{1/2}.
\label{PorbRL}
\end{equation}
Hence the orbital periods of the upper ($P_{\rmn{u}}$) and lower
($P_{\ell}$) edges of the period gap relate to the corresponding
donor's radii at these locations, $R_{\rmn{L},\rmn{u}}$ and
$R_{\rmn{L},\ell}$ as
\begin{equation}
\frac{P_{\rmn{u}}}{P_{\rmn{\ell}}}=\left(\frac{R_{\rmn{L,u}}}{R_{\rmn{L,\ell}}}\right)^{3/2}.
\label{PuPl}
\end{equation}
There is no term in $M_{2}$ as the mass of the secondary star remains
constant across the period gap. Given that
$P_{\rmn{u}}/P_{\rmn{\ell}}=3/2$, then
$R_{\rmn{L,u}}/R_{\rmn{L,\ell}}\equiv{f_{\rmn{conv}}}\approx{1.3}$.

\subsubsection{Mass Transfer in CVs}

The instantaneous mass transfer rate, $\dot{M}_{2}$, from the CV donor
with radius $R_{2}$ and Roche lobe radius $R_{2,\rmn{L}}$ is given by
\citep{ritter88}
\begin{equation}
-\dot{M}_{2}\approx\dot{M}_{0}\,\rmn{exp}\left(\frac{R_{2,\rmn{L}}-R_{2}}{H}\right),
\label{instdMdt}
\end{equation}
where $H\approx{10^{-4}R_{2}}$ is the photospheric scale height of the
donor star and $\dot{M}_{0}\approx{10^{-8}}$ $\rmn{M_{\odot}}$
yr$^{-1}$. We assume steady state mass transfer,
i.e. $\ddot{M}_{2}=0$. From equation (\ref{instdMdt}), this assumption
implies that the donor radius and Roche lobe radius move in step
\begin{equation}
\frac{\rmn{d}}{\rmn{dt}}(R_{2,\rmn{L}}-R_{2})=0,
\label{in_step}
\end{equation}
where we have also assumed that $R_{2}-R_{2,\rmn{L}}\ll{R_{2}}$.

We calculate the mass transfer rate for a given binary configuration
time step by considering the relative difference between the donor
radius and Roche lobe radius $\delta=(R_{2}-R_{2,\rmn{L}})/R_{2}$ from
BiSEPS as a function of trial values of the mass transfer rate. We
locate the root of the function $\delta(\dot{M}_{2})$ by employing a
bisection algorithm, and therefore obtaining the value of
$\dot{M}_{2}$. In practice we iterate until we find a value of
$\dot{M}_{2}$ where $|\delta|{\la{10^{-5}}}$. This procedure is
similar to that used by Kalogera et al. (2003), although our method
has been developed independently.

The bisection algorithm encounters numerical problems for those CVs
that are close to thermal or dynamical instability, or where the
secondary star has evolved significantly from the zero-age main
sequence (typically where $M_{2}/M_{\odot}\ga{1.0}$). Specifically,
the algorithm obtains unphysically high mass transfer rates at the
onset of mass transfer. Consequently, the evolutionary tracks of these
systems across the period gap are not calculated.

We estimate the present day number of dCVs that these systems would
produce for each model by multiplying the mean dCV lifetime with the
formation rate of these numerically unstable CVs (we apply equation
(\ref{N_dCV}) described in Section 3.1). The mean dCV lifetime,
$\langle\tau_{\rmn{dCV}}\rangle$, was calculated by evolving a typical
CV with a 0.4 M$_{\odot}$ donor and a 0.6 M$_{\odot}$ white
dwarf. These typical values were determined by calculating the
probability distribution of all zero-age CVs. We find that
$\langle\tau_{\rmn{dCV}}\rangle\approx{1053}$ Myr.

The estimated present day number of these dCVs are added onto the
present day numbers calculated from the population synthesis. The
fraction of the total dCV model population that these numerically
unstable CVs contribute are as follows. For the Hurley et al. (2002)
prescription of magnetic braking, the contribution is $\la{8.0}$ per
cent. For the Rappaport et al. (1983) prescription, the contribution
is $\la{20.0}$ per cent when $\gamma=2$ and $\la{30.0}$ per cent when
$\gamma=4$.

%We estimate the number of present day dCVs that these systems would
%produce for each model by applying equation (\ref{N_dCV}), which is
%described in Section 3.1. We determine the average value of the period
%gap crossing time in equation (\ref{N_dCV}),
%$\langle\tau_{\rmn{dCV}}\rangle$, by evolving a typical CV with a 0.4
%M$_{\odot}$ donor star and a 0.6 M$_{\odot}$ white dwarf. These
%typical values were obtained by calculating the probability
%distribution of all zero-age CVs in the $M_{1}-P_{\rmn{orb}}$ and
%$M_{2}-P_{\rmn{orb}}$ plane. We find that
%$\langle\tau_{\rmn{dCV}}\rangle\approx{1053}$ Myr. We then multiply
%this value of $\langle\tau_{\rmn{dCV}}\rangle$ to the formation rates
%of these numerically unstable CVs, which are calculated for each
%model.

%These estimated values of the present day number of dCVs that these
%numerically unstable CVs produce are added onto the present day
%numbers calculated from the population synthesis code. The percentage
%of the total dCV population that these numerically unstable CVs
%contribute are as follows. For the Hurley et al (2002) prescription of
%magnetic braking, the contribution is $\la{8.0}$ per cent. For the
%Rappaport et al (1983) prescription, the contribution is $\la{20.0}$
%per cent where $\gamma=2$ and $\la{30.0}$ per cent where $\gamma=4$.

\subsubsection{The Reaction of the Donor to Mass Loss}

We apply a simple analytical scheme that mimics the reaction of the
donor star to mass loss. The expansion of a low-mass donor as a result
of mass loss is a well established result from full stellar evolution
calculations and analytical considerations (Stehle et al. 1996). We
define the radius excess, $f$, as the factor by which the star has
exceeded its equilibrium radius, $R_{2,\rmn{eq}}$, due to mass
loss. We parameterise the evolution of $f$ according to a power law
expression
\begin{equation}
%f\equiv{\frac{R_{2,\rmn{CV}}}{R_{2,\rmn{eq}}}}=CM_{2}^{\Delta{\zeta}}\label{f},
f\equiv{\frac{R_{2,\rmn{CV}}}{R_{2,\rmn{eq}}}}=AM_{2}^{\beta}\label{f},
\end{equation}
%where the constants $\Delta{\zeta}\equiv{\zeta_{2}-\zeta_{1}}$ and $C=B/A$.
where $A$ and $\beta$ are constants.

At the upper edge of the period gap, the donor star exceeds its
equilibrium radius by a factor $f=f_{\rmn{conv}}=1.3$. At the point
where mass transfer just commences at $M_{2}=M_{2,\rmn{0}}$, we
require that the donor star is in thermal equilibrium
i.e. $f=f_{0}=1$. Thus we have

\begin{equation}
f_{0}=AM_{2,0}^{\beta},\label{f0}
\end{equation}
and
\begin{equation}
f_{\rmn{conv}}=AM_{2,\rmn{conv}}^{\beta},\label{fconv}
\end{equation}
Combining (\ref{f}), (\ref{f0}) and (\ref{fconv}) yields
\begin{equation}
f=f_{\rmn{conv}}\left(\frac{M_{2}}{M_{2,\rmn{conv}}}\right)^{\beta},\label{ffull}
\end{equation}
where
\begin{equation}
\beta=\frac{\rmn{ln}(f_{0}/f_{\rmn{conv}})}{\rmn{ln}(M_{2,0}/M_{2,\rmn{conv}})}.\label{beta}
\end{equation}

\begin{figure}
  \begin{center}
%  \vspace{174pt}
  \includegraphics[scale=0.45]{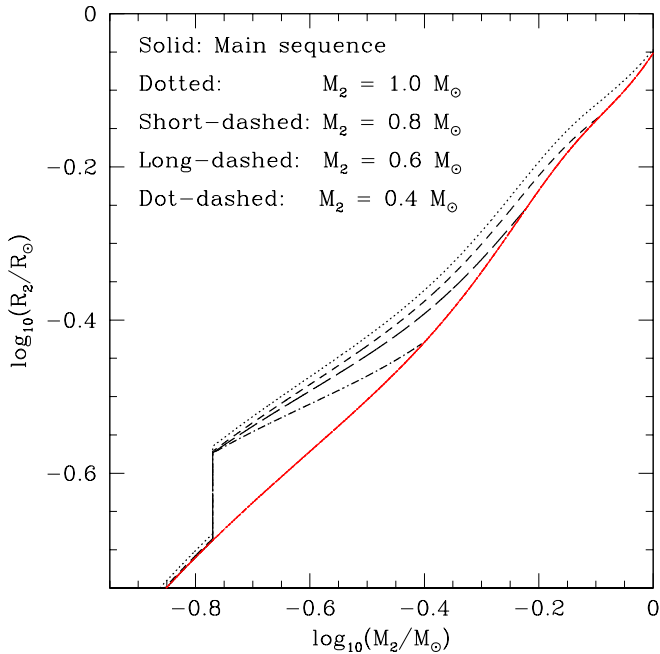}
  \end{center}
  \caption{Mass-radius relations for CVs with different
  initial donor masses.}
  \label{tracks}
\end{figure}

The value of the donor's radius calculated by BiSEPS is then
multiplied by $f$. We assume furthermore that the donor's effective
temperature, $T_{\rmn{eff}}$, is the same as its isolated, main
sequence equivalent (Stehle et al. 1996). The luminosity, $L_{2}$,
of the donor star is calculated from the Stefan-Boltzmann Law
\begin{equation}
L_{2}=4\pi{R_{2}}^{2}\sigma{T_{\rmn{eff}}^{4}}.
\label{SBLaw}
\end{equation}
where $\sigma$ is Stefan-Boltzmann constant. Hence $L_{2}$ will be
larger by a factor of $f^{2}$ compared to models without mass
loss. The mass-radius relation of several CV evolutionary sequences
calculated with this prescription is shown in Figure \ref{tracks}.

In order for the lower edge of the period gap to be in the correct
location of $P_{\ell}=2.0$ h, we disrupt magnetic braking when
$M_{2}=M_{\rmn{2,conv}}=0.17$ $\rmn{M_{\odot}}$.

Based on homology relations, Stehle et al. (1996) derived a first
order differential equation that describes how the donor star's radius
reacts to mass loss. This scheme shows that CVs with different initial
system parameters at the onset of mass transfer quickly converge to a
single, uniform evolutionary track, consistent with the well defined
position of the period gap.

Our initial scheme to describe the reaction of the donor to mass loss
in terms of the mass loss timescale $\tau_{\dot{M}_{2}}$ and its
Kelvin-Helmholtz timescale $\tau_{\rmn{KH}}$, was closely based on
that of Stehle et al. (1996), according to
\begin{equation}
f=\kappa\frac{\tau_{\rmn{KH}}}{\tau_{\dot{M}_{2}}}+1,
\label{f_tscls}
\end{equation}
where the value of the constant $\kappa$ can be determined by
stipulating that $f=f_{\rmn{conv}}=1.3$ at the upper edge of the
period gap. However, we encountered numerical instabilities due to
feedback between the donor radius and the mass transfer rate. The
scheme we use instead does not reproduce convergence of CV evolution
onto a single track, but does ensure that the donors do exceed their
equilibrium radius by the same amount by the time they become fully
convective. As such, we ensure that we reproduce the observed width of
the period gap with a well defined location. As we are interested in
the evolution of systems across the period gap, rather than the
detailed CV evolution above the gap, we believe the prescription we
adopt is satisfactory.

Following \citet{kk95} once the donor star detaches from its Roche
lobe it re-attains its equilibrium radius on the timescale
\begin{equation}
\tau=\left(1-\frac{1}{f_{\rmn{conv}}}\right)\frac{D}{\zeta_{\rmn{eff}}-\zeta_{\rmn{ad}}}\tau_{J},\label{tau}
\end{equation}
where $\zeta_{\rmn{eff}}=\rmn{dln}R_{2}/\rmn{dln}M_{2}$ is the
effective mass-radius index, $\zeta_{\rmn{ad}}$ is the adiabatic
mass-radius exponent, $\tau_{J}$ is the angular momentum loss rate
timescale and the stability denominator $D\approx{1}$ is given by
equation (16) in \citet{kk95}. These constants are evaluated just
before the CV becomes detached at the upper edge of the period
gap. Typically, $\tau\approx{10}^{7}$ yr. In the detached phase the
radius excess follows the relation
\begin{equation}
f(t)=1-(1-f_{\rmn{conv}})e^{-t/\tau},\label{f(t)}
\end{equation}
where $t$ is the time since the donor detached from its Roche lobe.

\section{Results and Analysis}

We now present the results of our population synthesis
calculations. These are summarised in Table \ref{table01}, which lists
the present day population of dCVs, gPCEBs and the ratio dCV:gPCEB in
the period gap. For completeness we also list the total present day
formation rates of all PCEBs, and CVs above and below the period
gap. In all our models, we also consider different forms of the IMRD,
with $\nu=-0.99$, 0.0 or 1.0. We also determine the secondary mass
from the same IMF as the primary star. Our reference IMRD is $n(q)=1$.

\begin{figure}
  \begin{center}
%  \vspace{174pt}
  \includegraphics[scale=0.4]{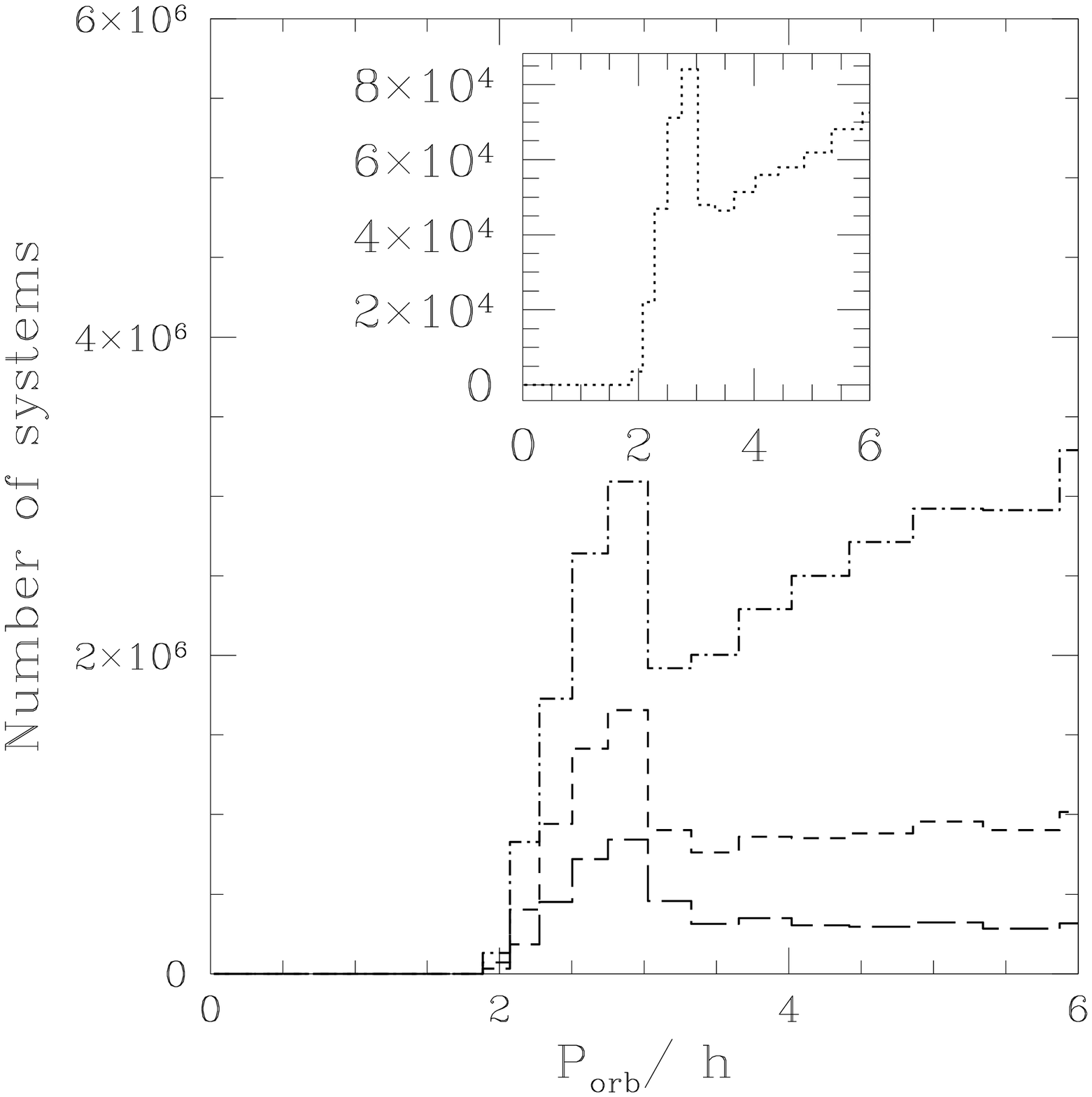}
  \includegraphics[scale=0.4]{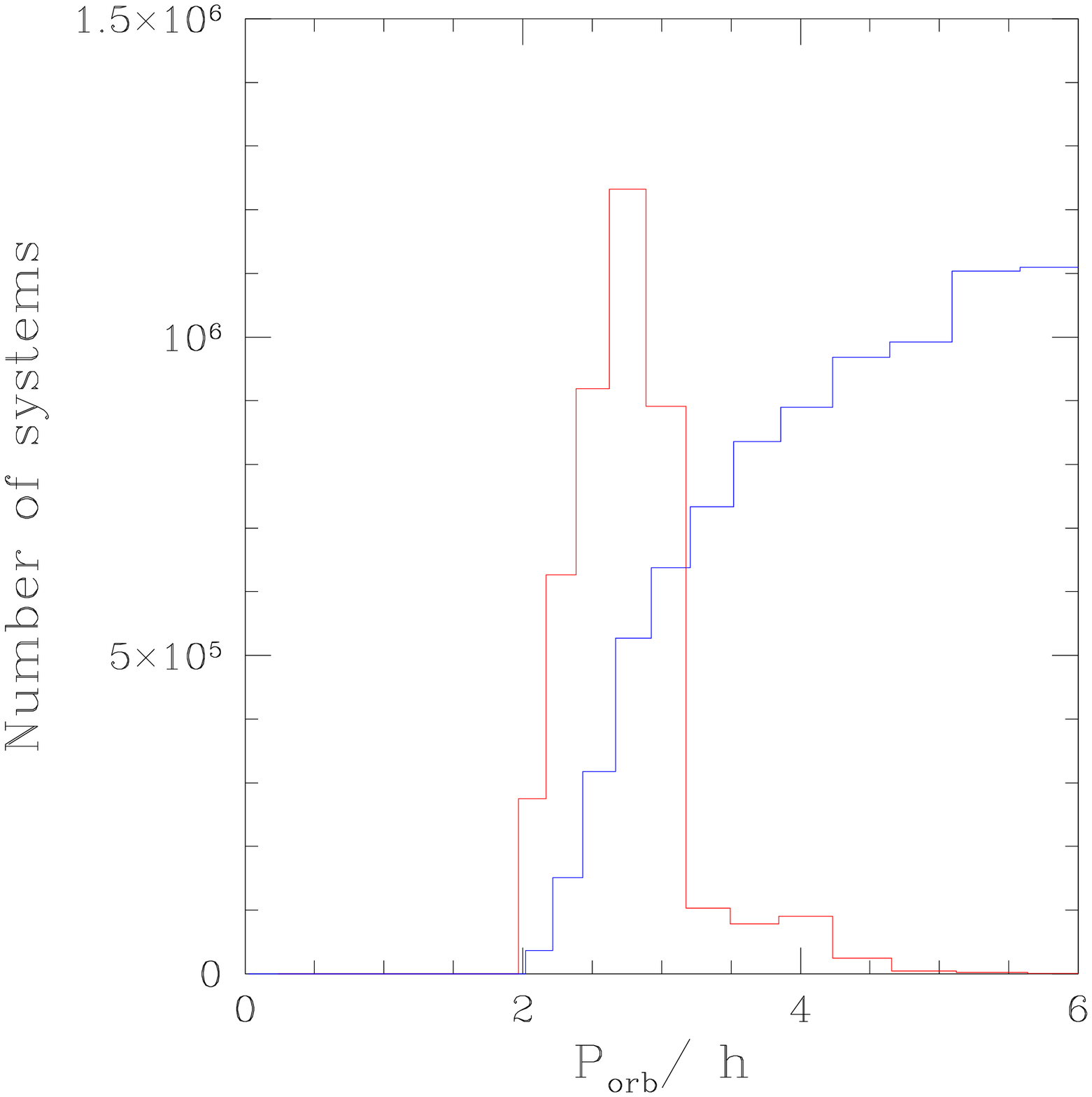}
  \end{center}
  \caption{Top panel: orbital period distribution of dCVs and gPCEBs
  combined for our reference model hA, and for different initial mass
  ratio distributions. Long-dashed: $n(q)\propto{q}$; short-dashed:
  $n(q)=1$ (our reference IMRD); dot-dashed: where $M_{2}$ is picked
  independently from the same IMF as $M_{1}$ according to equation
  (\ref{M1dist}); dotted: $n(q)\propto{q}^{-0.99}$ (inset). The peak
  apparent in all the distributions is due to the population of dCVs;
  the gPCEB orbital period distributions are increasing monotonically
  with $P_{\rmn{orb}}$. Bottom panel: orbital period distribution of
  dCVs (red) and gPCEBs (blue) for our reference model hA, and for
  $n(q)=1$.}
  \label{Pdist_HA_01}
\end{figure}

The excess of dCVs over gPCEBs is illustrated in the top panel of
Figure \ref{Pdist_HA_01}; a prominent peak within the period gap in
the orbital period distributions of the combined dCV and gPCEB
populations. These have been calculated from our reference model hA,
for different assumptions on the initial distribution of the secondary
mass. The cut-off in the number of systems at 2.0 hours is because the
smallest secondary mass considered in both the dCV and gPCEB systems
is 0.17 M$_{\odot}$, as such systems become semi-detached at 2.0
hours. The bottom panel of Figure \ref{Pdist_HA_01} shows the separate
orbital period distributions of dCVs (red) and gPCEBs (blue) for our
reference model hA, and for $n(q)=1$. The ratio dCV:gPCEB within the
period gap gives an indication of the height of the peak.

%\begin{figure}
%  \begin{center}
%    \includegraphics[scale=0.4]{figY.eps}
%  \end{center}
%  \caption{Orbital period distribution of dCVs (red) and gPCEBs (blue)
%  for our reference model hA, and for $n(q)=1$.}
%  \label{Pdist_HA_02}
%\end{figure}

\subsection{Varying $\alpha_{\rmn{CE}}$: Models hA to hCE5}

We begin by examining how the present day populations of dCVs and
gPCEBs within the period gap are affected by varying the value of the
global parameter $\alpha_{\rmn{CE}}$ for $n(q)=1$ (Figure \ref{num_alpha}).
An understanding of these trends will yield an understanding of the overall
trend of dCV:gPCEB with $\alpha_{\rmn{CE}}$.

\subsubsection{The gPCEB Population}

%\begin{figure*}
%  \centering
%  \begin{minipage}{175mm}
%    \includegraphics[scale=0.55]{fig5a.eps}
%    \includegraphics[scale=0.55]{fig5b.eps}
%    \caption{From left to right: the present day population of gPCEBs
%    in the period gap (blue) and dCVs (red) as a function of
%    $\alpha_{\rmn{CE}}$, and the ratio dCV:gPCEB as a function of
%    $\alpha_{\rmn{CE}}$. Long-dashed: $n(q)\propto{q}$; short-dashed:
%    $n(q)=1$ (our reference IMRD); dot-dashed: where $M_{2}$ is picked
%    independently from the same IMF as $M_{1}$ according to equation
%    (\ref{M1dist}); dotted: $n(q)\propto{q}^{-0.99}$ (insets).}
%    \label{num_alpha}
%  \end{minipage}
%\end{figure*}

\begin{figure*}
  \centering
  \begin{minipage}{175mm}
    \includegraphics[scale=0.55]{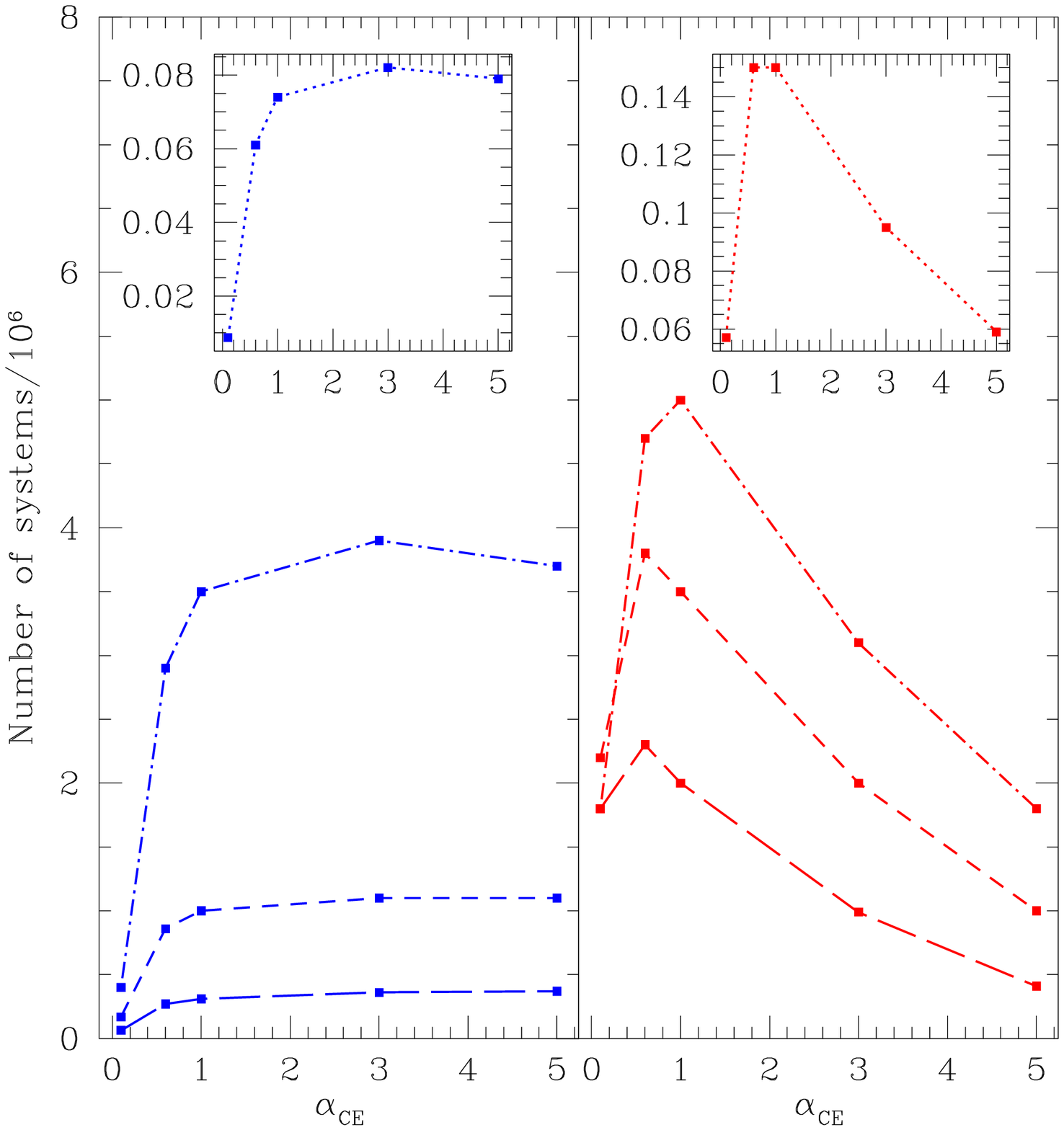}
    \includegraphics[scale=0.55]{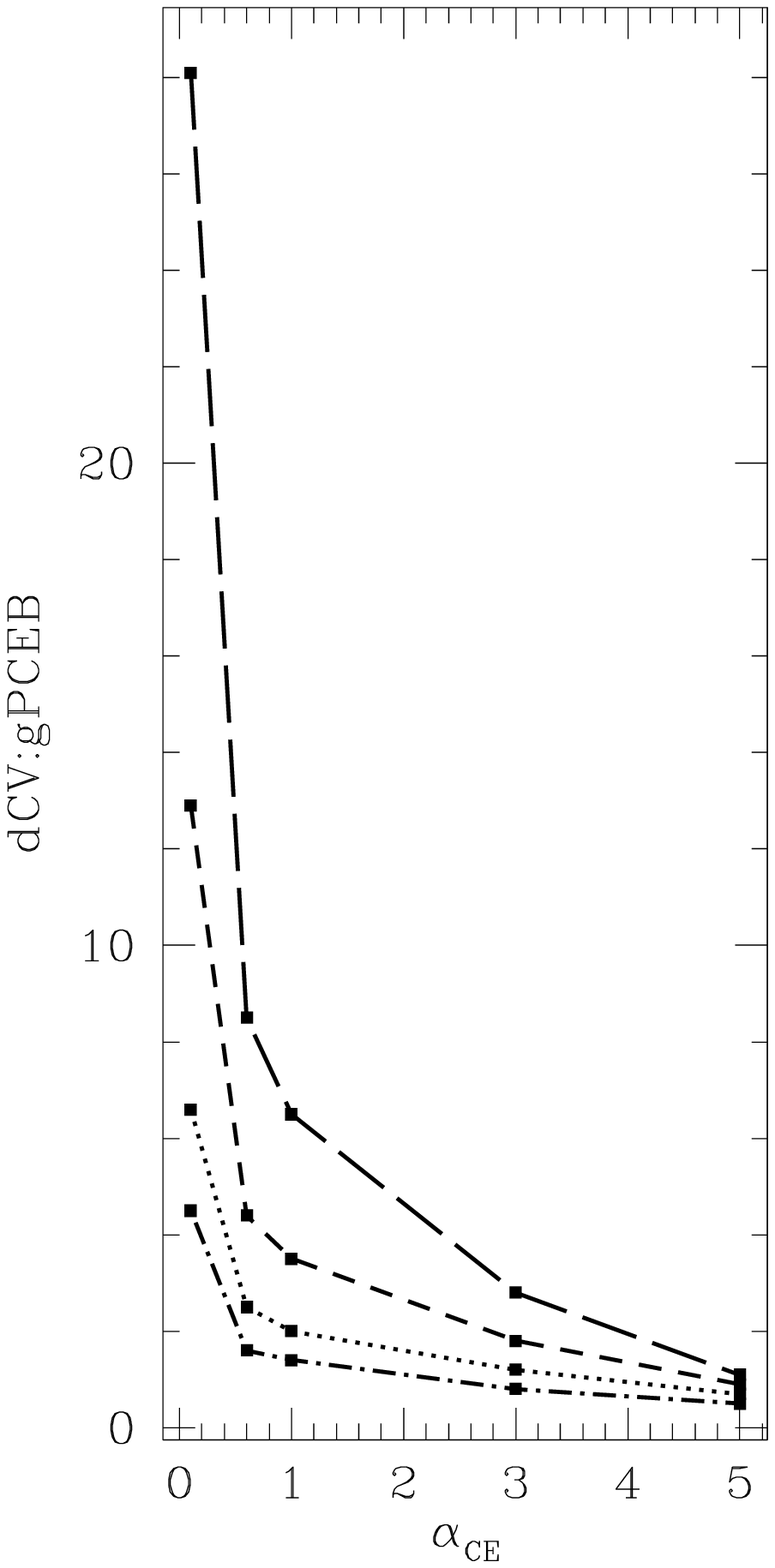}
    \caption{From left to right: the present day population of gPCEBs
    in the period gap (blue) and dCVs (red) as a function of
    $\alpha_{\rmn{CE}}$, and the ratio dCV:gPCEB as a function of
    $\alpha_{\rmn{CE}}$. Long-dashed: $n(q)\propto{q}$; short-dashed:
    $n(q)=1$ (our reference IMRD); dot-dashed: where $M_{2}$ is picked
    independently from the same IMF as $M_{1}$ according to equation
    (\ref{M1dist}); dotted: $n(q)\propto{q}^{-0.99}$ (insets).}
    \label{num_alpha}
  \end{minipage}
\end{figure*}

The left panel of Figure \ref{num_alpha} shows that increasing the
value of $\alpha_{\rmn{CE}}$ from 0.1 to 1.0 increases the present day
number of gPCEBs within the period gap (from $1.7\times{10}^{5}$ to
$1.0\times{10}^{6}$ for our reference IMRD). This levels off (here to
$1.1\times{10}^{6}$) for $\alpha_{\rmn{CE}}\ga{1.0}$.

\begin{figure}
  \includegraphics[scale=0.4]{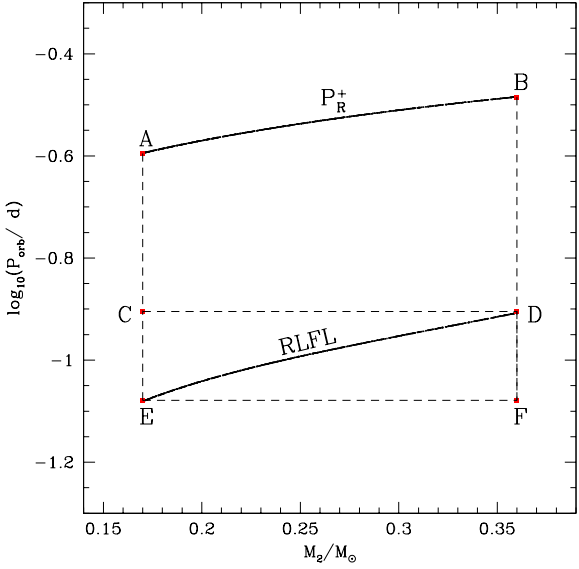}
  \caption{The dashed box, $CDEF$, shows the region $\Re$ defined by
  $0.17\le{M_{2}/\rmn{M}_{\odot}}\le{0.36}$ and
  $\rmn{log}_{10}(P_{\ell})\le{\rmn{log}_{10}(P_{\rmn{orb}}/\rmn{d})}\le\rmn{log}_{10}(P_{\rmn{u}})$. The
  gPCEBs within $\Re$ will fill their Roche lobes within the period
  gap. The largest orbital period a gPCEB can have that will evolve
  into the period gap within the lifetime of the Galaxy is given by
  the boundary $P^{+}_{\Re}$. Thus the total area from which gPCEBs
  can flow into the period gap within the Galactic lifetime is the
  region $ABCD$.}
  \label{gPCEB_boundaries}
\end{figure}

To understand this consider Figure \ref{gPCEB_boundaries}, where the
dashed box $CDEF$ defined by $0.17\le{M_{2}/\rmn{M_{\odot}}}\le{0.36}$
and
$\rmn{log}_{10}\,(P_{\ell})\le{P_{\rmn{orb}}/\rmn{h}}\le{\rmn{log}_{10}\,(P_{\rmn{u}})}$
shows the location of gPCEBs in the period gap. We designate this
region as $\Re$. Not only will $\Re$ consist of gPCEBs that formed
there, but also of gPCEBs that evolve into $\Re$ from longer orbital
periods. If $t_{\rmn{u}}=10$ Gyr is the maximum time allowed for the
gPCEB to reach the upper edge of the period gap and $t_{\rmn{GR}}$ is
the AM loss timescale, $J/\dot{J}$, due to gravitational radiation,
then the largest orbital period from which a gPCEB can evolve into the
period gap within the lifetime of the Galaxy, $P^{+}_{\Re}$, can be
solved from
%The longest orbital period from which a gPCEB can evolve into
%$\Re$ within the lifetime of the Galaxy is shown by the boundary
%$P^{+}_{\Re}$. If $t_{\rmn{u}}=10$ Gyr is the maximum time allowed for
%the gPCEB to reach the upper edge of the period gap and $t_{\rmn{GR}}$
%is the AM loss timescale due to gravitational radiation then}
\begin{equation}
t_{\rmn{u}}=\frac{t_{\rmn{GR}}(P^{+}_{\Re})-t_{\rmn{GR}}(P_{\rmn{u}})}{8},
\label{tu}
\end{equation}
(e.g. Kolb \& Stehle 1996). The location of $P^{+}_{\Re}$ is also
shown in Figure \ref{gPCEB_boundaries}. Therefore, the total area from
which gPCEBs can evolve into $\Re$ within the Galactic lifetime is
given by the region $ABCD$.

%In the figure this is approximated by the location of
%systems with a constant AM loss timescale of 10 Gyr due to
%gravitational radiation. Therefore, the total area from which gPCEBs
%can evolve into $\Re$ within the lifetime of the Galaxy is given by
%the region $ABCD$.}

\begin{figure}
  \includegraphics[scale=0.4]{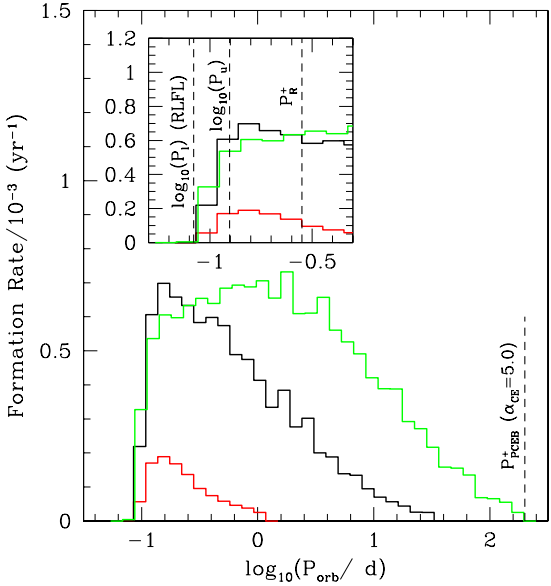}
  \caption{The formation rate of gPCEBs as a function of
  $\rmn{log}_{10}\,(P_{\rmn{orb}}/ \rmn{d})$ for $n(q)=1$ and for
  $\alpha_{\rmn{CE}}=0.1$ (red), 1.0 (black) and 5.0 (green). Shown in
  the inset are the period gap boundaries log$_{10}$$P_{\ell}$ (which
  is also the RLFL for 0.17 M$_{\odot}$ secondaries) and
  log$_{10}$$P_{\rmn{u}}$, and $P^{+}_{\Re}$. The upper boundary
  $P^{+}_{\rmn{PCEB}}$ for $\alpha_{\rmn{CE}}=5.0$ is shown in the
  main plot as the dashed line. Note that $P^{+}_{\rmn{PCEB}}$, and in
  fact the whole distribution function, shifts to shorter periods as
  $\alpha_{\rmn{CE}}$ decreases.}
  \label{gPCEB_PreCV_Pops1}
\end{figure}

The whole population of gPCEBs is also bound within an upper limit,
$P^{+}_{\rmn{PCEB}}$ (shown in Figure \ref{gPCEB_PreCV_Pops1}), and
the Roche lobe-filling limit (RLFL) which is the orbital period where
systems become semi-detached. As the value of $\alpha_{\rmn{CE}}$ is
decreased the boundary $P^{+}_{\rmn{PCEB}}$ and the population bound
within will shift to shorter orbital periods. This is because, for low
ejection efficiencies, more orbital energy is required to eject the
envelope from the system. The resulting PCEB (and therefore gPCEB)
population will lie at shorter orbital periods.

%\begin{figure}
%  \includegraphics[scale=0.4]{fig6.eps}
 % \caption{The dashed box, $CDEF$, shows the region $\Re$ defined by
%  $0.17\le{M_{2}/\rmn{M}_{\odot}}\le{0.36}$ and
%  $\rmn{log}_{10}(P_{\ell})\le{\rmn{log}_{10}(P_{\rmn{orb}}/\rmn{d})}\le\rmn{log}_{10}(P_{\rmn{u}})$. The
%  gPCEBs within $\Re$ will fill their Roche lobes within the period
%  gap. The largest orbital period a gPCEB can have that will evolve
%  into the period gap within the lifetime of the Galaxy is given by
%  the boundary $P^{+}_{\Re}$. Thus the total area from which gPCEBs
%  can flow into the period gap within the Galactic lifetime is the
%  region $ABCD$.}
%  \label{gPCEB_boundaries}
%\end{figure}

%The upper boundary arises from the
%most massive progenitor primary (which can produce a white dwarf) just
%filling its Roche lobe at the tip of the asymptotic giant branch.

%The left panel of Figure \ref{gPCEB_boundaries}
%shows the location of $P^{+}_{\rmn{PCEB}}$ for
%$\alpha_{\rmn{CE}}=1.0$, while the right panel shows its location for
%$\alpha_{\rmn{CE}}=0.1$, which lies at shorter orbital periods. This
%is because, for low ejection efficiencies, more orbital energy is
%required to eject the envelope from the system. The resulting PCEB
%population therefore lies at shorter orbital periods.}

We find that the flux of gPCEBs from $ABCD$ into $\Re$ is larger than
the birthrate of gPCEBs within $\Re$ by a factor of $\sim{7}$. If the
typical lifetime of a gPCEB within $\Re$ is
$\langle\tau_{\rmn{gPCEB}}\rangle\approx{380}$ Myr and the flux of
gPCEBs from $ABCD$ into $\Re$ is $F_{\Re}$ then the present day number
of gPCEBs within the period gap, $N_{\rmn{gPCEB}}$, can be
approximated as
\begin{equation}
N_{\rmn{gPCEB}}\approx{F_{\Re}}\times{\langle\tau_{\rmn{gPCEB}}\rangle}.
\label{N_gPCEB}
\end{equation}
Clearly $F_{\Re}$ will depend on the formation rate of gPCEBs in
$ABCD$. Figure \ref{gPCEB_PreCV_Pops1} shows the formation rate of
gPCEBs as a function of $\rmn{log}_{10}\,(P_{\rmn{orb}}/\rmn{d})$ for
$n(q)=1$, and for $\alpha_{\rmn{CE}}=0.1$ (red), 1.0 (black) and 5.0
(green). The distribution function is flat-topped and tails off
towards longer orbital periods. As $\alpha_{\rmn{CE}}$ increases and
the whole population of gPCEBs shifts to longer orbital periods, the
formation rate of the gPCEBs between log$_{10}(P_{\rmn{u}})$ and
$P^{+}_{\Re}$ (inset)- and therefore in $ABCD$- changes as a
consequence; at its smallest for $\alpha_{\rmn{CE}}=0.1$ but with
little change for $\alpha_{\rmn{CE}}\ga{1.0}$.

From equation (\ref{N_gPCEB}) the trend in $N_{\rmn{gPCEB}}$ with
$\alpha_{\rmn{CE}}$ will therefore mirror that of the birthrate of
gPCEBs in $ABCD$. This is indeed the case as shown in the left panel
of Figure \ref{num_alpha}.

\begin{table*}
  \centering
  \begin{minipage}{175mm}
    \caption{Present day population of
    dCVs, gPCEBs and the ratio dCV:gPCEB in the period gap for various
    values of $\alpha_{\rmn{CE}}$ and various forms of magnetic
    braking. Also quoted is the present day formation rate of all PCEBs,
    and of CVs above and below the upper edge of the period gap.}
    \begin{tabular}{@{}lllcccc@{}}
%      \hline
      \hline
      Model  &  Number of          &  Number of         &  Formation rate of  &  Formation rate   &  Formation rate  & dCV:gPCEB \\
             &    dCVs            &    gPCEBs          &  PCEBs (yr$^{-1}$)  & of CVs $\ge$ 3.0 h & of CVs $<$ 3.0 h  &   \\
             &                    &                    &                     &    (yr$^{-1}$)     &  (yr$^{-1}$) &\\
      \hline
      \hline
       $n(q)\propto{q^{-0.99}}$, $0<q\le{1}$\\
      \hline
      \hline
      hCE01  & $5.7\times{10}^{4}$ & $8.7\times{10}^{3}$   & $2.1\times{10}^{-4}$ & $6.7\times{10}^{-5}$ & $3.7\times{10}^{-5}$ & 6.6 \\
      hCE06  & $1.5\times{10}^{5}$ & $6.1\times{10}^{4}$   & $1.5\times{10}^{-3}$ & $1.8\times{10}^{-4}$ & $2.3\times{10}^{-4}$ & 2.5 \\
      hA     & $1.5\times{10}^{5}$ & $7.4\times{10}^{4}$   & $2.0\times{10}^{-3}$ & $1.8\times{10}^{-4}$ & $2.7\times{10}^{-4}$ & 2.0 \\
      hCE3   & $9.5\times{10}^{4}$ & $8.2\times{10}^{4}$   & $3.1\times{10}^{-3}$ & $1.1\times{10}^{-4}$ & $2.4\times{10}^{-4}$ & 1.2 \\
      hCE5   & $5.9\times{10}^{4}$ & $7.9\times{10}^{4}$   & $3.5\times{10}^{-3}$ & $7.7\times{10}^{-5}$ & $2.0\times{10}^{-4}$ & 0.7 \\
      \hline
      hPWR05 & $1.6\times{10}^{5}$ & $5.2\times{10}^{4}$   & $1.6\times{10}^{-3}$ & $1.8\times{10}^{-4}$ & $1.9\times{10}^{-4}$ & 3.1 \\
      hPWR1  & $1.4\times{10}^{5}$ & $2.5\times{10}^{4}$   & $1.3\times{10}^{-3}$ & $1.7\times{10}^{-4}$ & $1.0\times{10}^{-4}$ & 5.6 \\
      hPWR2  & $1.0\times{10}^{5}$ & $3.3\times{10}^{3}$   & $1.1\times{10}^{-3}$ & $1.4\times{10}^{-4}$ & $2.2\times{10}^{-5}$ & 30.3  \\
      \hline
      rvj2A  & $2.2\times{10}^{5}$ & $7.4\times{10}^{4}$   & $2.0\times{10}^{-3}$ & $2.5\times{10}^{-4}$ & $2.8\times{10}^{-4}$ & 3.0 \\
      rvj4A  & $2.4\times{10}^{5}$ & $7.4\times{10}^{4}$   & $2.0\times{10}^{-3}$ & $3.2\times{10}^{-4}$ & $2.8\times{10}^{-4}$ & 3.2 \\
      \hline
      \hline
      $n(q)=1.0$, $0<q\le{1}$ (our reference IMRD)\\
      \hline
      \hline
      hCE01  &  $2.2\times{10}^{6}$  & $1.7\times{10}^{5}$  & $9.8\times{10}^{-3}$ & $3.1\times{10}^{-3}$ & $7.0\times{10}^{-4}$ &  12.9  \\
      hCE06  &  $3.8\times{10}^{6}$  & $8.6\times{10}^{5}$  & $5.5\times{10}^{-2}$ & $5.6\times{10}^{-3}$ & $3.3\times{10}^{-3}$ &  4.4   \\
      hA     &  $3.5\times{10}^{6}$  & $1.0\times{10}^{6}$  & $7.2\times{10}^{-2}$ & $5.0\times{10}^{-3}$ & $3.3\times{10}^{-3}$ &  3.5 \\
      hCE3   &  $2.0\times{10}^{6}$  & $1.1\times{10}^{6}$  & $1.0\times{10}^{-1}$ & $2.7\times{10}^{-3}$ & $2.5\times{10}^{-3}$ &  1.8   \\
      hCE5   &  $1.0\times{10}^{6}$  & $1.1\times{10}^{6}$  & $1.2\times{10}^{-1}$ & $1.5\times{10}^{-3}$ & $1.9\times{10}^{-3}$ &  0.9   \\
      \hline
      hPWR05 &  $4.1\times{10}^{6}$  & $7.7\times{10}^{5}$  & $6.8\times{10}^{-2}$ & $5.2\times{10}^{-3}$ & $2.8\times{10}^{-3}$ &  5.3   \\
      hPWR1  &  $3.9\times{10}^{6}$  & $4.4\times{10}^{5}$  & $6.5\times{10}^{-2}$ & $5.2\times{10}^{-3}$ & $1.9\times{10}^{-3}$ &  8.9  \\
      hPWR2  &  $3.1\times{10}^{6}$  & $7.2\times{10}^{4}$  & $6.1\times{10}^{-2}$ & $4.8\times{10}^{-3}$ & $6.8\times{10}^{-4}$ &  43.1  \\
      \hline
      rvj2A  &  $5.5\times{10}^{6}$  & $1.0\times{10}^{6}$  & $7.2\times{10}^{-2}$ & $7.6\times{10}^{-3}$ & $3.6\times{10}^{-3}$ &  5.5   \\
      rvj4A  &  $6.0\times{10}^{6}$  & $1.0\times{10}^{6}$  & $7.2\times{10}^{-2}$ & $1.0\times{10}^{-2}$ & $3.9\times{10}^{-3}$ &  6.0   \\
      \hline
      \hline
      $n(q)\propto{q}$, $0<q\le{1}$\\
      \hline
      \hline
      hCE01  & $1.8\times{10}^{6}$ & $6.4\times{10}^{4}$   & $1.1\times{10}^{-2}$ & $3.2\times{10}^{-3}$ & $3.1\times{10}^{-4}$ & 28.1 \\
      hCE06  & $2.3\times{10}^{6}$ & $2.7\times{10}^{5}$   & $5.8\times{10}^{-2}$ & $4.2\times{10}^{-3}$ & $1.4\times{10}^{-3}$ & 8.5  \\
      hA     & $2.0\times{10}^{6}$ & $3.1\times{10}^{5}$   & $7.5\times{10}^{-2}$ & $3.6\times{10}^{-3}$ & $1.1\times{10}^{-3}$ & 6.5  \\
      hCE3   & $9.9\times{10}^{5}$ & $3.6\times{10}^{5}$   & $1.1\times{10}^{-1}$ & $1.7\times{10}^{-3}$ & $6.1\times{10}^{-4}$ & 2.8  \\
      hCE5   & $4.1\times{10}^{5}$ & $3.7\times{10}^{5}$   & $1.2\times{10}^{-1}$ & $7.0\times{10}^{-4}$ & $4.5\times{10}^{-4}$ & 1.1  \\
      \hline
      hPWR05 & $2.3\times{10}^{6}$ & $2.5\times{10}^{5}$   & $7.7\times{10}^{-2}$ & $3.7\times{10}^{-3}$ & $1.1\times{10}^{-3}$ & 9.2 \\
      hPWR1  & $2.3\times{10}^{6}$ & $1.6\times{10}^{5}$   & $7.8\times{10}^{-2}$ & $3.7\times{10}^{-3}$ & $9.1\times{10}^{-4}$ & 14.4 \\
      hPWR2  & $2.0\times{10}^{6}$ & $3.2\times{10}^{4}$   & $7.8\times{10}^{-2}$ & $3.5\times{10}^{-3}$ & $5.3\times{10}^{-4}$ & 62.5 \\
      \hline
      rvj2A  & $3.2\times{10}^{6}$ & $3.1\times{10}^{5}$   & $7.5\times{10}^{-2}$ & $5.7\times{10}^{-3}$ & $1.5\times{10}^{-3}$ & 10.3   \\
      rvj4A  & $3.6\times{10}^{6}$ & $3.1\times{10}^{5}$   & $7.5\times{10}^{-2}$ & $8.9\times{10}^{-3}$ & $1.9\times{10}^{-3}$ & 11.6   \\
      \hline
      \hline
      $M_{2}$ from IMF according to eqn. (\ref{M1dist})\\
      \hline
      \hline
      hCE01  & $1.8\times{10}^{6}$ & $4.0\times{10}^{5}$  & $6.5\times{10}^{-3}$ & $1.9\times{10}^{-3}$ & $1.7\times{10}^{-3}$ & 4.5  \\
      hCE06  & $4.7\times{10}^{6}$ & $2.9\times{10}^{6}$  & $4.7\times{10}^{-2}$ & $4.8\times{10}^{-3}$ & $1.1\times{10}^{-2}$ & 1.6  \\
      hA     & $5.0\times{10}^{6}$ & $3.5\times{10}^{6}$  & $6.5\times{10}^{-2}$ & $4.8\times{10}^{-3}$ & $1.3\times{10}^{-2}$ & 1.4  \\
      hCE3   & $3.1\times{10}^{6}$ & $3.9\times{10}^{6}$  & $1.0\times{10}^{-1}$ & $3.1\times{10}^{-3}$ & $1.2\times{10}^{-2}$ & 0.8  \\
      hCE5   & $1.8\times{10}^{6}$ & $3.7\times{10}^{6}$  & $1.2\times{10}^{-1}$ & $2.1\times{10}^{-3}$ & $1.0\times{10}^{-2}$ & 0.5  \\
      \hline
      hPWR05 & $5.2\times{10}^{6}$ & $2.4\times{10}^{6}$  & $4.4\times{10}^{-2}$ & $4.9\times{10}^{-3}$ & $9.0\times{10}^{-3}$ & 2.2  \\
      hPWR1  & $4.7\times{10}^{6}$ & $1.2\times{10}^{6}$  & $2.9\times{10}^{-2}$ & $4.6\times{10}^{-3}$ & $4.3\times{10}^{-3}$ & 3.9  \\
      hPWR2  & $3.1\times{10}^{6}$ & $1.4\times{10}^{5}$  & $1.7\times{10}^{-2}$ & $3.6\times{10}^{-3}$ & $8.2\times{10}^{-4}$ & 22.1 \\
      \hline
      rvj2A  & $6.2\times{10}^{6}$ & $3.5\times{10}^{6}$  & $6.5\times{10}^{-2}$ & $6.3\times{10}^{-3}$ & $1.4\times{10}^{-2}$ & 1.8   \\
      rvj4A  & $6.8\times{10}^{6}$ & $3.5\times{10}^{6}$  & $6.5\times{10}^{-2}$ & $7.6\times{10}^{-3}$ & $1.4\times{10}^{-2}$ & 1.9   \\
      \hline
%      \hline
\label{table01}
\end{tabular}
\end{minipage}
%\label{table01}
\end{table*}

%\begin{figure}
%  \includegraphics[scale=0.4]{fig6.eps}
%  \caption{The dashed box, $CDEF$, shows the region $\Re$ defined by
%  $0.17\le{M_{2}/\rmn{M}_{\odot}}\le{0.36}$ and
%  $\rmn{log}_{10}(P_{\ell})\le{\rmn{log}_{10}(P_{\rmn{orb}}/\rmn{d})}\le\rmn{log}_{10}(P_{\rmn{u}})$. The
%  gPCEBs within $\Re$ will fill their Roche lobes within the period
%  gap. The largest orbital period a gPCEB can have that will evolve
%  into the period gap within the lifetime of the Galaxy is given by
%  the boundary $P^{+}_{\Re}$. Thus the total area from which gPCEBs
%  can flow into the period gap within the Galactic lifetime is the
%  region $ABCD$.}
%  \label{gPCEB_boundaries}
%\end{figure}

%\begin{figure}
%  \includegraphics[scale=0.4]{fig7.eps}
%  \caption{The formation rate of gPCEBs as a function of
%  $\rmn{log}_{10}\,(P_{\rmn{orb}}/ \rmn{d})$ for $n(q)=1$ and for
%  $\alpha_{\rmn{CE}}=0.1$ (red), 1.0 (black) and 5.0 (green). Shown in
%  the inset are the period gap boundaries log$_{10}$$P_{\ell}$ (which
%  is also the RLFL for 0.17 M$_{\odot}$ secondaries) and
%  log$_{10}$$P_{\rmn{u}}$, and $P^{+}_{\Re}$. The upper boundary
%  $P^{+}_{\rmn{PCEB}}$ for $\alpha_{\rmn{CE}}=5.0$ is shown in the
%  main plot as the dashed line. Note that $P^{+}_{\rmn{PCEB}}$, and in
%  fact the whole distribution function, shifts to shorter periods as
%  $\alpha_{\rmn{CE}}$ decreases.}
%  \label{gPCEB_PreCV_Pops1}
%\end{figure}

\subsubsection{The dCV Population}

As with the present day population of gPCEBs, there is also an
increase in the present day number of dCVs associated with an increase
in $\alpha_{\rmn{CE}}$ from 0.1 to 1.0, as shown in the middle panel
of Figure \ref{num_alpha}. For our reference IMRD, the number of dCVs
increases from $2.2\times{10}^{6}$ to $3.5\times{10}^{6}$. In contrast
to the gPCEB population however, the number of dCVs decreases with a
further increase in the value of $\alpha_{\rmn{CE}}$.

If the formation rate of dCVs is $B_{\rmn{dCV}}$ and their average
lifetime is $\langle\tau_{\rmn{dCV}}\rangle$, then the present-day
population of dCVs can be approximated by
\begin{equation}
N_{\rmn{dCV}}\approx{B_{\rmn{dCV}}}\times{\langle\tau_{\rmn{dCV}}\rangle}.
\label{N_dCV}
\end{equation}
If we now assume that the populations above the period gap of pre-CVs
(with $M_{2}>M_{\rmn{MS,conv}}$) and CVs are in a steady state, we
then have for their respective formation rates $B_{\rmn{preCV}}$ and
$B_{\rmn{CV}}$
\begin{equation}
B_{\rmn{PreCV}}\approx{B_{\rmn{CV}}}\approx{B_{\rmn{dCV}}}.
\label{B_dCV}
\end{equation}
The formation rate of dCVs is therefore linked to that of
these pre-CVs above the period gap.

%\begin{figure}
%  \includegraphics[scale=0.47,angle=90]{fig8.eps}
%  \caption{The formation rate of pre-CVs for $n(q)=1$ on the
%  $M_{2}-\rmn{log}_{10}(P_\rmn{orb})$ plane for
%  $\alpha_{\rmn{CE}}=0.1$ (top panel), 1.0 (middle panel) and 5.0
%  (bottom panel). Pre-CVs that lie above the solid line in the bottom
%  panel formed through a case C CE phase, while those that lie below
%  it formed through a case B CE phase with a naked helium star
%  remnant. The colour bar at the top indicates the formation rate
%  (Myr$^{-1}$) per bin area
%  $\rmn{d}M_{2}\,\rmn{d}\rmn{log}_{10}(P_{\rmn{orb}})$.}
%  \label{PreCV_BR_y2y3_1}
%\end{figure}

\begin{figure}
  \includegraphics[scale=0.47,angle=90]{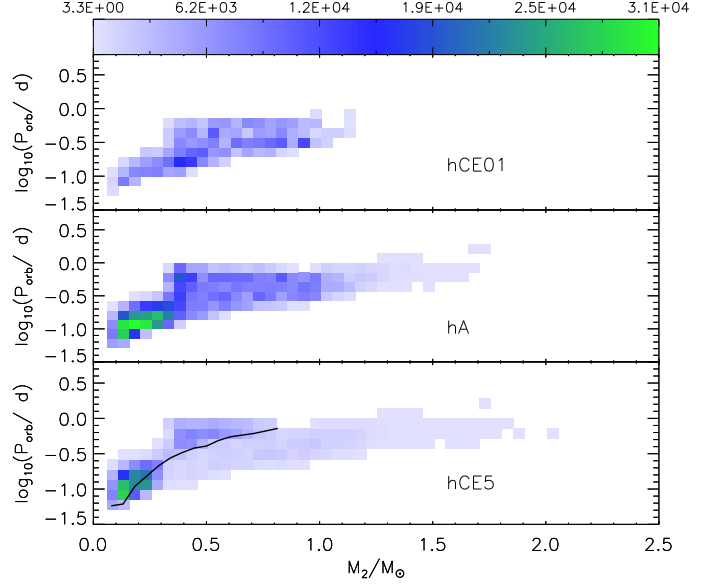}
  \caption{The formation rate of pre-CVs for $n(q)=1$ on the
  $M_{2}-\rmn{log}_{10}(P_\rmn{orb})$ plane for
  $\alpha_{\rmn{CE}}=0.1$ (top panel), 1.0 (middle panel) and 5.0
  (bottom panel). Pre-CVs that lie above the solid line in the bottom
  panel formed through a case C CE phase, while those that lie below
  it formed through a case B CE phase with a naked helium star
  remnant. The colour bar at the top indicates the formation rate
  (Myr$^{-1}$) per bin area
  $\rmn{d}M_{2}\,\rmn{d}\rmn{log}_{10}(P_{\rmn{orb}})$.}
  \label{PreCV_BR_y2y3_1}
\end{figure}

Figure \ref{PreCV_BR_y2y3_1} shows the present-day formation
rate of pre-CVs for $n(q)=1$ on the $M_{2}-\rmn{log_{10}}(P_{\rmn{orb}})$
plane, for $\alpha_{\rmn{CE}}=0.1$ (top panel), 1.0 (middle panel) and
5.0 (bottom panel), where the colour bar at the top indicates the
formation rate in Myr$^{-1}$ per bin area. The lower boundary in each
of the populations is the RLFL, while the upper boundary is given by
those PCEBs where the time taken for the primary progenitor to fill
its Roche lobe plus the time taken for the secondary to subsequently
come into contact with its Roche lobe is 10 Gyr (i.e. the Galactic
lifetime). The shape of the upper boundary is a consequence of the AM
loss mechanism; gravitational radiation for systems with
$M_{2}\le{M_{\rmn{MS,conv}}}$ or a combination of magnetic braking and
gravitational radiation for systems with $M_{2}>M_{\rmn{MS,conv}}$.

Inspecting the upper and middle panels of Figure \ref{PreCV_BR_y2y3_1}
shows that there is an increase in the formation rate of pre-CVs from
$\alpha_{\rmn{CE}}=0.1$ to $\alpha_{\rmn{CE}}=1.0$. This is a
consequence of the whole population of PCEBs being shifted to longer
orbital periods for larger ejection efficiencies, and hence more
systems surviving the CE phase. For $\alpha_{\rmn{CE}}=5.0$ in the
lower panel, however, there is an overall decrease in the formation
rate of pre-CVs. Progenitor systems with even shorter initial orbital
periods than those for models hCE01 or hA can now survive the CE
phase. Such progenitors result in pre-CVs that formed through a case B
CE phase with a naked helium star remnant. These systems lie below the
solid line in the bottom panel of Figure \ref{PreCV_BR_y2y3_1}, while
those above the line are those pre-CVs that formed through a case C CE
phase. The latter systems have ended up at much longer orbital periods
than for models hCE01 and hA. The solid line represents those pre-CVs
whose progenitor primaries just filled their Roche lobes on the base
of the AGB. It is the disappearance of case C remnants and the
appearance of the less abundant case B remnants that causes the
decrease in the total pre-CV formation rate.
 
From equations (\ref{N_dCV}) to (\ref{B_dCV}) this trend in
$B_{\rmn{preCV}}$ should be mirrored in the present day number of
dCVs, which is the case as shown in the middle panel of Figure
\ref{num_alpha}.
%\textbf{The right panel of Figure \ref{gPCEB_PreCV_Pops1} shows the
%formation rate of pre-CVs as a function of
%$\rmn{log}_{10}\,(P_{\rmn{orb}}/ \rmn{d})$ for $n(q)=1$, and for
%$\alpha_{\rmn{CE}}=0.1$ (red), 1.0 (black) and 5.0 (green). The upper
%limit, $P^{+}_{\rmn{PreCV}}$, is the longest orbital period a PCEB can
%have that will become semi-detached within the lifetime of the
%Galaxy. In the figure this is approximated, as before, by the location of
%systems with a constant AM loss timescale of 10 Gyr, here due to
%a combination of magnetic braking (according to equation
%(\ref{hurley_mb})) and gravitational radiation. Also shown is the
%lower limit, $P_{\rmn{u}}$, which is also the RLFL for systems with
%$M_{2}=M_{\rmn{MS,conv}}$.}

%\textbf{As $\alpha_{\rmn{CE}}$ increases the whole population of PCEBs
%shifts to longer orbital periods causing the formation rate to change
%as a consequence; $B_{\rmn{PreCV}}$ increases for
%$\alpha_{\rmn{CE}}=0.1$ to 1.0, but then decreases again for
%$\alpha_{\rmn{CE}}$=5.0. This trend is indeed mirrored in
%$N_{\rmn{dCV}}$, as shown in the middle panel of Figure
%\ref{num_alpha}.}

\subsubsection{Overall Trends}

These trends in the populations of gPCEBs and dCVs are the same for
all forms of the IMRD function, and where the secondary mass is
determined independently from the IMF according to (\ref{M1dist}). The
largest population of gPCEBs and dCVs is obtained when the secondary
mass is determined for this latter case. The smallest population
numbers are found for $n(q)\propto{q}^{-0.99}$.

Combining the trends of dCVs and gPCEBs gives a decrease in the ratio
of dCV:gPCEB with increasing $\alpha_{\rmn{CE}}$, seen in the
right-most panel of Figure \ref{num_alpha}. For our reference IMRD
function, dCV:gPCEB decreases from 12.9 for $\alpha_{\rmn{CE}}=0.1$ to
0.9 for $\alpha_{\rmn{CE}}=5.0$. Again this trend is the same for all
forms of the IMRD function. We note that the ratio dCV:gPCEB decreases
with decreasing value of $\nu$. To understand why, note that the dCVs
will have formed from CVs filling their Roche lobes at long orbital
periods above the period gap. This is in contrast to gPCEBs in the
period gap which have secondary masses in the range
$0.17\le{M_{2}/M_{\odot}}\le{0.36}$. Thus, for a given primary mass,
the dCV progenitor systems will have, on average, more massive
secondaries than found in gPCEBs. The IMRD $n(q)\propto{q}$ will
mostly favour the formation of systems with more massive secondaries,
thus mostly favouring the formation of dCV progenitors over the gPCEB
progenitors. Thus we obtain the largest values of dCV:gPCEB. This is
in contrast with the IMRD of the form $n(q)\propto{q}^{-0.99}$ which
most favour the formation of systems with less massive secondaries,
thereby mostly favouring the formation of gPCEB progenitors over dCV
progenitors. We therefore obtain the smallest values of dCV:gPCEB out
of the three forms of the IMRD. The smallest values of dCV:gPCEB are
obtained when the secondary mass is determined independently from the
primary mass.

\subsection{$\alpha_{\rmn{CE}}$ as a function of Secondary Mass: models hPWR05 to hPWR2}

\begin{figure*}
  \centering
  \begin{minipage}{175mm}
    \includegraphics[scale=0.55]{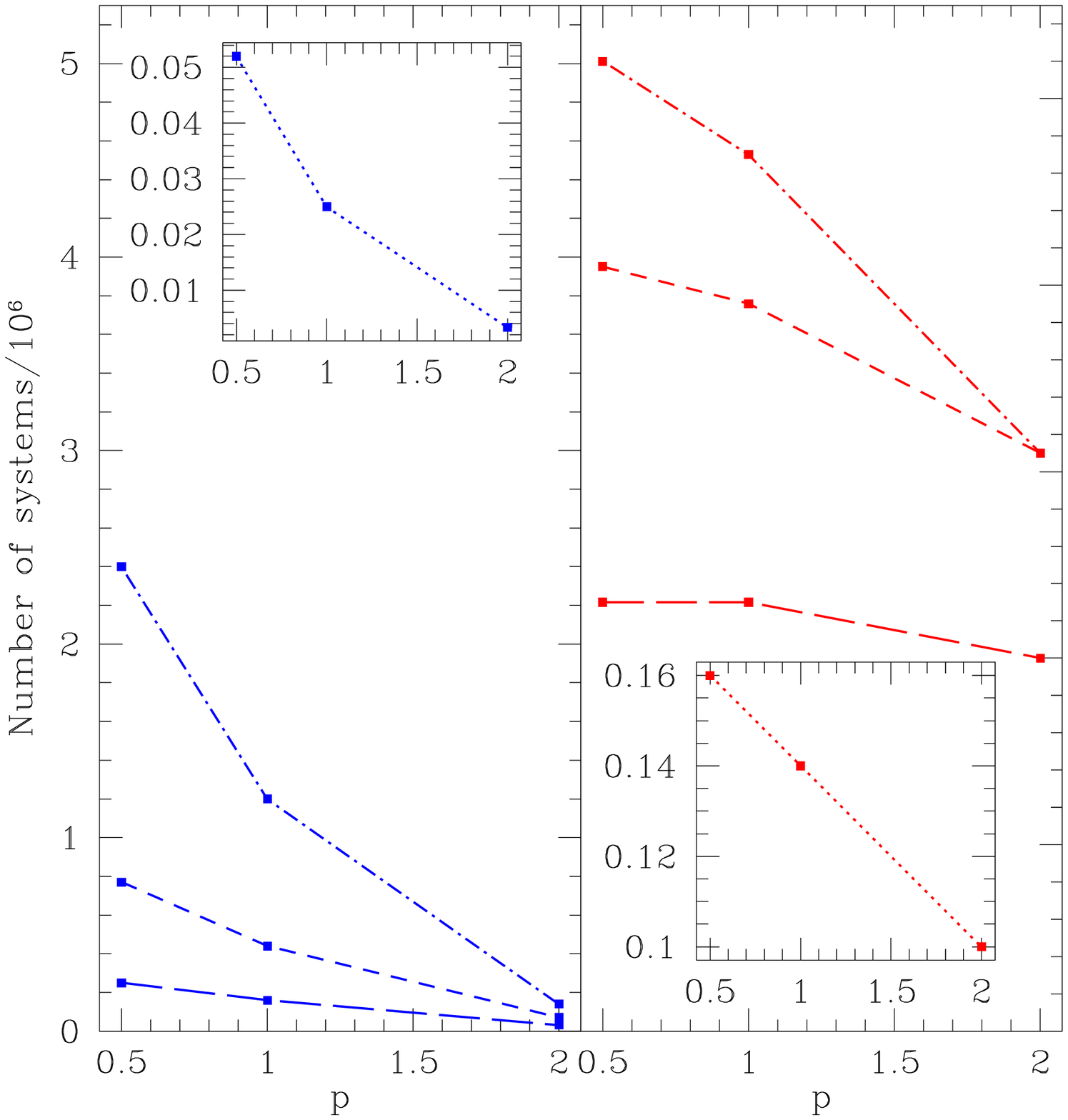}
    \includegraphics[scale=0.55]{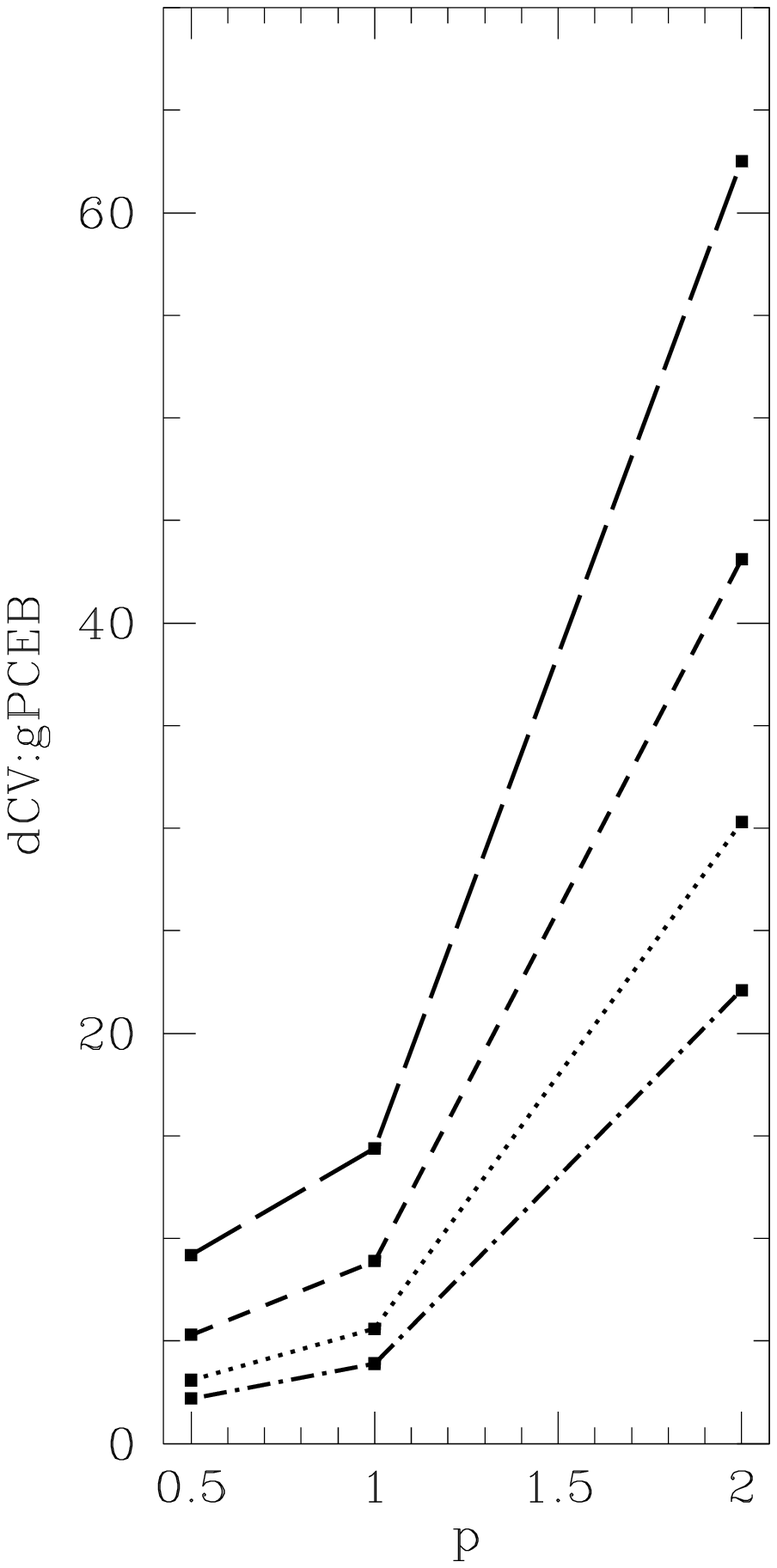}
    \caption{ From left to right: the present day population of gPCEBs
    in the period gap (blue) and dCVs (red) as a function of the power
    index $p$, and the ratio dCV:gPCEB as a function of
    $p$. Long-dashed: $n(q)\propto{q}$; short-dashed: $n(q)=1$;
    dot-dashed: where $M_{2}$ is picked independently from the same
    IMF as $M_{1}$ according to equation (\ref{M1dist}); dotted:
    $n(q)\propto{q}^{-0.99}$ (insets).}
    \label{num_n}
  \end{minipage}
\end{figure*}

In models where $\alpha_{\rmn{CE}}$ is determined from equation
(\ref{alpha1}) both the present day populations of gPCEBs and dCVs
decrease as the power index $p$ is increased. For our reference IMRD
function, the left panel of Figure \ref{num_n} shows that the number
of gPCEBs decreases from $7.7\times{10}^{5}$ to
$7.2\times{10}^{4}$. The middle panel shows that the number of dCVs
decreases slightly from $4.1\times{10}^{6}$ to $3.1\times{10}^{6}$. As
with the population of gPCEBs for models hA to hCE5, this trend
depends on the flow rate of gPCEBs into the region $\Re$, and how this
changes as the whole population of PCEBs is shifted with
$\alpha_{\rmn{CE}}$.

\subsubsection{The gPCEB Population}

The dark-grey area in Figure \ref{alpha_m2} indicates the range of
secondary masses for the gPCEB systems. As $p$ increases from 0.5 to
2, the average corresponding value of $\alpha_{\rmn{CE}}$ for these
systems decreases from $\sim{0.5}$ to $\sim{0.1}$. As a consequence
the population of gPCEBs will shift to shorter orbital periods with
increasing $p$. This in turn will change the formation rate of gPCEBs
in the region $ABCD$ and hence the flux $F_{\Re}$.

As can be seen in the left panel of Figure \ref{num_alpha}, a value of
$\alpha_{\rmn{CE}}$ decreasing from $\sim{0.5}$ to $\sim{0.1}$ leads
to the decrease in the present-day number of gPCEBs in the period
gap. This corresponds to the decrease in the present-day number of
gPCEBs with increasing $p$ seen in the left panel of Figure
\ref{num_n}.

\subsubsection{The dCV Population}

As with the dCVs for models hA to hCE5, the population of dCVs for
hPL05 to hPL2 depends on the formation rate of pre-CVs with
$M_{2}>M_{\rmn{MS,conv}}$ above the period gap. The light-grey region
in Figure \ref{alpha_m2} shows the range of secondary masses for dCV
progenitor systems. For this range in secondary mass there is, on
average, little change in $\alpha_{\rmn{CE}}$. Indeed, the dominant
secondary mass group for pre-CV progenitors is $M_{2}\la{0.6}$
M$_{\odot}$; here an increase in $p$ from 0.5 to 2 corresponds to a
decrease in $\alpha_{\rmn{CE}}$ from $\sim{0.7}$ to $\sim{0.3}$. Hence
from the middle panel of Figure \ref{num_alpha} a slight decrease in
$N_{\rmn{dCV}}$ can be expected. This is indeed the case as shown in
the middle panel of Figure \ref{num_n}.

\subsubsection{Overall Trends}

The trend in the present day populations of gPCEBs and dCVs with $p$
is the same for all forms of the IMRD function, and where the
secondary mass is determined from the same IMF as the primary mass. As
before, the largest population of dCVs and gPCEBs occurs when the
secondary mass is determined from the same IMF as the primary. The
smallest populations occur when $n(q)\propto{q}^{-0.99}$.

As a consequence of the behaviour discussed in Sections 3.2.1 and
3.2.2, the ratio dCV:gPCEB increases as $p$ increases. For our
reference model, dCV:gPCEB increases from 5.3 for $p=0.5$ to 43.1 for
$p=2$. The variation of dCV:gPCEB with $p$ is shown in the most right
hand panel in Figure \ref{num_n}.

%The ratio dCV:gPCEB increases with
%increasing $p$ because the the decrease $N_{\rmn{dCV}}$ with
%increasing $p$ is more modest than the decrease in $N_{\rmn{gPCEB}}$
%with $p$. 

Note that these ratios are very large for $p=2$ as PCEBs with low-mass
donors will encounter a merger. Hence the vast majority of all CVs
will form above the period gap, with very few gPCEBs within the period
gap. Politano \& Weiler (2007) suggested that such a strong dependence
of $\alpha_{\rmn{CE}}$ on $M_{2}$ could explain the lack of CVs with
sub-stellar secondaries.  However, we note that a lack of calculated
gPCEBs with low-mass secondaries appears to be inconsistent with
observations (Ritter \& Kolb 2003, catalogue Edition 7.7 (2006)),
which show a modest number of PCEBs with $M_{2}\approx{0.17}$
M$_{\odot}$. We therefore believe that the $p=2$ model is not a
realistic one. As with models hA to hCE5, dCV:gPCEB decreases with
decreasing values of $\nu$. The reasons for this are the same as those
described in Section 3.1. The smallest value of dCV:gPCEB occurs when
the secondary mass is determined independently from the primary mass.

%\begin{figure}
%  \includegraphics[scale=0.4]{gPCEB.birth.hPL05_hPL1_hPL2.y3.eps}
%  \caption{\textbf{The formation rate of gPCEBs as a function of
%  $\rmn{log}_{10}\,(P_{\rmn{orb}}/\rmn{d})$ for $n(q)=1$ and for
%  $p=0.5$ (red), 1.0 (black) and 2.0 (green). Shown in the inset by
%  the dashed lines are the period gap boundaries log$_{10}$$P_{\ell}$
%  (which is also the RLFL for 0.17 M$_{\odot}$ secondaries) and
%  log$_{10}$$P_{\rmn{u}}$, and $P^{+}_{\Re}$. The upper boundary
%  $P^{+}_{\rmn{PCEB}}$ for $p=0.5$ is shown in the main plot (dashed
%  line). Note that $P^{+}_{\rmn{PCEB}}$ shifts to shorter values as
%  $p$ increases.}}
%  \label{gPCEB_PreCV_Pops2}
%\end{figure}

%\begin{figure}
%  \includegraphics[scale=0.47,angle=90]{Hurley_MB_ClassicPreCV_PCEB_zahst_hPL05_hPL1_hPL2_ctu_s0.00_ty2y3_output.eps}
%  \caption{\textbf{The formation rate of pre-CVs for $n(q)=1$ on the
%  $M_{2}-\rmn{log}_{10}(P_{\rmn{orb}})$ plane for $p=0.5$ (top panel),
%  1.0 (middle panel) and 2.0 (bottom panel). The colour bar shows the
%  formation rate (Myr$^{-1}$) per bin area
%  $\rmn{d}M_{2}\,\rmn{d}\rmn{log}_{10}(P_{\rmn{orb}})$.}}
%  \label{PreCV_BR_y2y3_2}
%\end{figure}

\subsection{Alternative Magnetic Braking Laws: Models rvj2A and rvj4A}

We now consider the impact of different forms of magnetic braking on
gPCEBs and dCVs. Increasing $\gamma$ from 2 to 4 increases the present
day number of dCVs from $5.5\times{10}^{6}$ for $\gamma=2$ to
$6.0\times{10}^{6}$ for $\gamma=4$, for our reference IMRD. The
Rappaport et al. (1983) prescription of magnetic braking gives a
larger present day population of dCVs than the Hurley et al. (2002)
prescription, between any IMRD function; for our reference model hA,
with $n(q)=1$, the present day number of dCVs is $3.5\times{10}^{6}$.

The pre-CV progenitor systems of CVs that form above the period gap
will initially be driven by magnetic braking. Increasing the strength
of magnetic braking will shift the upper limit $P_{\rmn{preCV}}^{+}$
to longer orbital periods for systems with
$M_{2}>M_{\rmn{MS,conv}}$. Thus more systems will become semi-detached
within the lifetime of the Galaxy. From equations (\ref{N_dCV}) to
(\ref{B_dCV}), this will result in an increase in the formation rate
of pre-CVs above the period gap, and therefore increase the formation
rate of dCVs. In turn, this will increase the present day population
of dCVs.

Note that the number of present day gPCEBs does not change with
magnetic braking law. In our model, the mass at which isolated stars
become fully convective is 0.35 $\rmn{M_{\odot}}$ (Hurley et
al. 2002). Thus, the evolution of gPCEBs with $M_{2}\le{0.35}$
M$_{\odot}$ will be driven by gravitational radiation only. As we are
only considering PCEBs with secondaries in the range
$0.17\le{M_{2}}/M_{\odot}\le{0.36}$, only very few systems will evolve
via magnetic braking while the larger majority will be driven via
gravitational radiation.

These trends in the population of dCVs and gPCEBs result in an
increase in dCV:gPCEB with increasing magnetic braking strength. For
our reference IMRD function, and where $\alpha_{\rmn{CE}}=1.0$, the
ratio increases from 3.5 for the Hurley et al. (2002) form of magnetic
braking, to 6.0 for the Rappaport et al. (1983) prescription,
$\gamma=4$. The largest values of dCV:gPCEB occur when the IMRD
function is $n(q)\propto{q}$, while the smallest occur when
$n(q)\propto{q}^{-0.99}$.

\section{Discussion}

\subsection{Observational Predictions}

In this section we consider the number of present day dCVs and gPCEBs,
and hence the ratio dCV:gPCEB, given the commonly assumed form of the
IMRD and values of $\alpha_{\rmn{CE}}$. We can then gauge whether such
an excess of dCVs over gPCEBs can be determined observationally.

\begin{figure}
  \centering
    \includegraphics[scale=0.4]{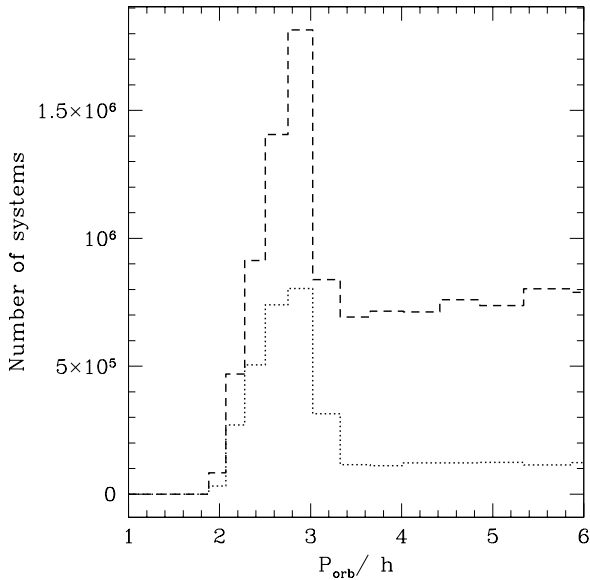}
    \caption{The orbital period distribution of the combined
    population of dCVs and gPCEBs for $n(q)=1$,
    $\alpha_{\rmn{CE}}=0.6$ (dashed line) and $\alpha_{\rmn{CE}}=0.1$
    (dotted line).}
    \label{y3_hCE01-hCE06}
\end{figure}

Numerical calculations of the CE phase suggest that
$\alpha_{\rmn{CE}}=0.13$ to 0.6 \citep{il93}, while the IMRD seems to
be flat in $q$, i.e. $n(q)=1$ (Goldberg, Mazeh \& Latham 2003; Mazeh
et al. 1992; Duquennoy \& Mayor 1991). From Table \ref{table01} we can
estimate that the number of dCVs currently occupying the Galaxy is
$(2.2-3.8)\times{10}^{6}$, while the number of gPCEBs is in the range
$(1.7-8.6)\times{10}^{5}$, giving a factor of $\sim{4}$ to $\sim{13}$
more dCVs than gPCEBs. Hence we predict a prominent peak in the
orbital period distribution of short orbital period WDMS systems as
shown in Figure \ref{y3_hCE01-hCE06}. In determining these
predictions, we have assumed that magnetic braking takes the form
shown in equation (\ref{hurley_mb}).

Of course, the AM loss rate associated with magnetic braking is not
very well constrained, and indeed it has been suggested that the
magnetic braking strength is $\approx{100}$ weaker than the forms we
have used in this investigation (Andronov et al. 2003). However, we
have not considered the forms of magnetic braking that include a
saturation point if the secondary star has a spin frequency larger
than some critical value $\Omega_{\rmn{crit}}$ (Sills et al. 2000;
Ivanova \& Taam 2003).This is because, in the case of the Sills et
al. (2000) form of magnetic braking, the AM loss rate would be too
small to sufficiently drive the donor star out of thermal
equilibrium, and consequently produce a period gap. Furthermore, the
Sills et al. (2000) form of magnetic braking predict mass transfer
rates at the upper edge of the period gap that are too low compared to
observations, and as such would not explain the presence of nova-like
CVs above the period gap (e.g. Kolb 2002). The Ivanova \& Taam (2003)
form of magnetic braking on the other hand, which is $\sim$ 2 to 3
times weaker than the Verbunt \& Zwaan (1982) form, would drive the
donor star out of thermal equilibrium but not enough to explain the
observed width of the period gap.

Such forms of magnetic braking are therefore inconsistent with the
disrupted magnetic braking model. Instead, alternative mechanisms must
exist that combine in strength to reach the value that we consider at
the upper edge of the period gap.

\subsection{The Location of the Period Gap}

We now discuss the impact the location of the period gap has on our
results. We note that Knigge (2006), using a similar method to ours,
determined that the period gap edges are 2.15 h and 3.18 h, which
agrees with our determined period gap width of one hour. We attribute
the slight difference between the two locations of the period gap to
the two different catalogues used. Knigge (2006) used the Ritter \&
Kolb (2003) catalogue (Edition 7.6), while we use the latest Edition
7.7.

If the period gap was located according to Knigge (2006), this would
have the effect of increasing the population of gPCEBs within the
period gap, and consequently decrease the ratio dCV:gPCEB. The higher
values of $P_{\rmn{u}}$ and $P_{\ell}$ would shift the mass range of
gPCEBs to $0.19\le{M_{2}/\rmn{M}_{\odot}}\le{0.40}$, and therefore a
larger fraction of gPCEBs (those with
$0.35<M_{2}/\rmn{M}_{\odot}\le{0.40}$) would be driven by magnetic
braking. In turn, the flow rate of gPCEBs from longer orbital periods
into $\Re$ would increase. We also point out that a similar
uncertainty is introduced with the value of $M_{\rmn{MS,conv}}$, as
this is not well constrained by stellar evolution models.

In their studies on circumbinary disk driven CVs Willems et al. (2005)
considered the CV population below the gap, while Willems et
al. (2007) studied the population above the gap.  These authors never
attempted to systematically model, or to achieve a best fit model for,
the combined population. This is why their adopted period gap
boundaries of 2.25 and 2.75 hours, respectively, result in an
implausibly narrow gap when taken together.

Nonetheless we investigated how a narrower period gap would affect our
results by calculating a further population model with the gap
boundaries set to 2.25 and 2.75 hours. We considered
$\alpha_{\rmn{CE}}=1.0$ with our standard IMRD, and re-calibrated
equation (\ref{hurley_mb}) to give the AM loss rate appropriate for
the narrower gap width. We find that dCV:gPCEB decreases to 2.1
compared to our model hA value of 3.5. Thus the prominence of the peak
would decrease somewhat. This is because the time for a dCV to cross a
narrower gap would decrease (we find
$\langle\tau_{\rmn{dCV}}\rangle\approx{438}$ Myr), decreasing the
present day population from $3.5\times{10}^{6}$ for model hA to
$1.1\times{10}^{6}$. The number of gPCEBs in the period gap also
decreases from $1.0\times{10}^{6}$ for model hA to $5.2\times{10}^{5}$
because the systems occupy a smaller area in period space.

We emphasise that a gap width of half an hour is a somewhat extreme
assumption. Despite this we still obtain a clear excess of dCVs over
gPCEBs in the period gap, and hence a prominent peak would still be
detected. Indeed, a prominent peak would still be detected for any
reasonable gap width.

\subsection{TTMT Systems}

Observational investigations have shown that in many CVs above the
period gap the donor is unlikely to be an unevolved main sequence
donor (e.g. Beuermann et al. 1998). Instead, the spectral types of the
donor are later than those of an isolated star of the same mass, while
donors at orbital periods longer than 5 to 6 hours are nuclear evolved
(Patterson et al. 2005; Knigge 2006). Theoretical calculations also
predict this effect (Baraffe \& Kolb 2000). It has been suggested that
such systems had an initial TTMT phase where the secondary star was
much more massive than the white dwarf (Schenker \& King 2002). Kolb
\& Willems (2005) find that as much as 40 per cent of zero-age CVs
form with a donor that has evolved more than half way through its main
sequence lifetime. G\"ansicke et al. (2003) have detected CNO
abundance anomalies consistent with nuclear evolved donors among
$\sim10-15$ per cent CVs in a sample observed in the ultraviolet,
corroborating the hypothesis that a significant fraction of the known
CV population may be post-TTMT systems.

Such CVs, if they appear after their initial TTMT phase above the
period gap, may contribute to the present day population of dCVs. Thus
our values of dCV:gPCEB, which are calculated from the evolution of
purely AM driven CVs, are lower limits. To calculate the contribution
of post-TTMT CVs to the total dCV population would require a
comprehensive treatment of the TTMT phase including the fate of the
transferred material, which determines the point of re-appearance of
the system as a `normal' AM driven CV.

\subsection{Magnetic CVs}

We now consider how the contribution of magnetic CVs (mCVs) may affect
the ratio dCV:gPCEB within the period gap. Intermediate polars (IPs)
are thought to evolve as non-magnetic CVs and hence become detached at
the upper edge of the period gap due to the disruption of magnetic
braking. Their contribution to the population of dCVs is therefore
already taken into account in our calculations, as are their
progenitors to the gPCEB population.

Polars, on the other hand, may not become detached at the upper edge
of the period gap, as it has been suggested that magnetic braking for
these systems is estimated to be 1 to 2 orders of magnitude weaker
than for non-magnetic CVs (Li, Wu \& Wickramasinghe 1994; Li, Wu \&
Wickramasinghe 1994b). This is supported observationally from their
white dwarf temperatures: the accretion-heated white dwarfs in polars
are consistently colder than those in non-magnetic CVs at similar
orbital periods, implying lower accretion rates, and hence angular
momentum loss rates (Araujo-Betancor et al. 2005). This difference is
especially pronounced above the period gap, where magnetic braking is
the dominant AM loss agent.

Furthermore, no progenitor WDMS systems of mCVs with a magnetic white
dwarf has as yet been detected (Liebert et al. 2005). It is possible
that such progenitors evolve in the same way as progenitors of
non-magnetic CVs. In such a case, the population of WDMS progenitors
of polars will already be taken into account in our calculations and
are included in the number of gPCEBs we obtained.

The overall effect of the above on our results would be to reduce the
number of dCVs and hence the ratio dCV:gPCEB by the fraction of polars
among the total CV population. From the Ritter \& Kolb (2002)
catalogue, Edition 7.7 (2006), we find that polars contribute
$\sim{13}$ per cent to the total CV population.

%Thus, polars may not contribute to the population of dCVs.

%As for the contribution of mCVs to the population of gPCEBs, Liebert
%at al. (2005) estimate that only 2 per cent of WDMS with a magnetic
%white dwarf contribute to the total PCEB population. Therefore, if
%they exist at all, WDMS systems with magnetic white dwarfs would
%contribute very little to the gPCEB population.

\subsection{PCEB Candidates}

Here we discuss the current situation regarding PCEB candidates and
the surveys presently searching for them. Schreiber \& G\"{a}nsicke
(2003) studied 30 well observed PCEBs, which were chosen to be
representative of PCEBs that will begin mass transfer within the
Hubble time. These systems have orbital periods $\la$2 days, have mass
ratios less than 1, and do not include secondaries that are
sub-giants. They found that the majority of systems contained young
($\la{5\times{10}^{8}}$ yrs), hot ($\approx$ 15 000-22 000 K) white
dwarfs. This is a consequence of the fact that until recently PCEB
candidates were almost exclusively identified in blue colour surveys
such as the Palomar Green survey. Thus, systems containing cool white
dwarfs and/or early type companion stars have been missed in most
previous work, with the exception of a few systems identified as
nearby large proper motion systems (e.g. RR\,Cae) or spectroscopic
binaries (e.g. V471\,Tau).

The potential of observational population studies of PCEBs has
improved over the past few years dramatically through the
SDSS. Because of the vast $ugriz$ colour space probed by SDSS, coupled
with the availability of high-quality follow-up spectroscopy of a
large number of objects with non-stellar colours, has already lead to
the identification of more than 1000 WDMS binaries (Silvestri et
al. 2007), most of them being by-products of the search for quasars. A
complementary program targeting specifically WDMS binaries containing
cool white dwarfs and/or early type companions is underway to
compensate the bias against such systems in the previous surveys
(Schreiber et al. 2007). For the first phase, the PCEBs among the full
sample of WDMS binaries have to be identified through radial velocity
studies. A first effort along these lines has been carried out by
Rebassa-Mansergas et al. (2007) who identified 18 PCEB candidates from
multiple SDSS spectroscopy obtained for 101 WDMS. Rebassa-Mansergas
(2008, in preparation) confirmed through follow-up spectroscopy so far
six of those candidates as PCEBs with orbital periods ranging from
164\,min to 1048\,min. Additional identification work of the PCEBs
among the SDSS WDMS is currently carried out at the VLT (Schreiber et
al. 2008) and the WHT (G\"ansicke et al. in preparation).

It is foreseeable that, sufficient observational effort being
invested, it will be possible to build up an orbital period
distribution of close WDMS binaries comprising potentially a few
hundred systems.  While the SDSS PCEB sample will not be free of
selection effects (see e.g. Rebassa-Mansergas et al. 2007; Pretorius,
Knigge \& Kolb 2006), those biases can be modelled to a large
degree. It appears hence feasible to subject the disrupted magnetic
braking model to a stringent test by comparing such an observed
orbital period distribution to the predictions of our work here
(Figure \ref{y3_hCE01-hCE06}).

\section{Conclusions}

We have performed population synthesis calculations to obtain the
present day population of two types of white dwarf-main sequence star
systems within the 2 to 3 hour cataclysmic variable period gap. The
first are post-CE binaries with secondaries that have
masses $0.17\le{M_{2}/\rmn{M_{\odot}}}\le{0.36}$, and so will commence
mass transfer in the period gap (gPCEBs). The second type are systems
that were CVs in the past, but detached at the upper edge of the
period gap as a consequence of disrupted magnetic braking, and are
crossing the period gap via gravitational radiation (dCVs).

Our calculations were repeated to consider constant, global values of
the CE ejection efficiency, $\alpha_{\rmn{CE}}$, and cases where
$\alpha_{\rmn{CE}}$ is a function of secondary mass according to
equation (\ref{alpha1}). We considered various forms of magnetic
braking according to equation (\ref{rvj_mb}) with $\gamma=2$ and 4,
and equation (\ref{hurley_mb}).

We find that there is a prominent peak in the orbital period
distribution of the combined dCV and gPCEB population, due to the
excess of dCVs over gPCEBs within the period gap. We find that the
ratio dCV:gPCEB, which gives an indication of the peak's height,
decreases with an increasing global value of $\alpha_{\rmn{CE}}$,
while increasing with increasing value of $p$. These trends are the
same for all assumptions on the initial secondary mass
distribution. The value of dCV:gPCEB ranges from 0.5 for model hCE5
where the secondary mass is determined independently from the primary
mass using the same IMF, to 62.5 for model hPWR2 where
$n(q)\propto{q}$.

%We find that the ratio dCV:gPCEB decreases with an increasing global
%value of $\alpha_{\rmn{CE}}$, while increasing with increasing value
%of $p$. These trends are the same for all assumptions on the initial
%secondary mass distribution. The value of dCV:gPCEB ranges from 0.5
%for model hCE5 where the secondary mass is determined independently
%from the primary mass using the same IMF, to 62.5 for model hPWR2
%where $n(q)\propto{q}$. The most likely value of dCV:gPCEB is between
%$\sim{4}$ to $\sim{13}$.

We find further that dCV:gPCEB increases with increasing magnetic
braking strength, although only slightly. For $n(q)=1$, dCV:gPCEB
increases from 3.5 for model hA (the weakest) to 6.0 for rvj4A (the
strongest).

For all our models we find that the values of dCV:gPCEB are largest
when the IMRD function has the form $n(q)\propto{q}$ i.e. on average
systems will have more massive secondary stars. The smallest values
occur when the secondary mass is determined from the same IMF as the
primary.

The most likely value of dCV:gPCEB is between $\sim{4}$ to $\sim{13}$,
thus we can expect a significant peak as shown in Figure
\ref{y3_hCE01-hCE06}.We suggest that if such a feature is observed in
the orbital period distribution of short orbital period WDMS binaries,
this would strongly corroborate the disruption of magnetic braking.

%We suggest that if an excess of dCVs over gPCEBs is determined
%observationally, this would strongly corroborate the disrupted
%magnetic braking model. Specifically, this excess would be revealed in
%an observationally determined orbital period distribution of all WDMS
%systems with orbital periods of a few hours.

\section*{Acknowledgements}

PJD acknowledges studentship support from the Science \& Technology
Facilities Council. BW acknowledges partial support from NASA BEFS
grant NNG06GH87G and BSF CAREER Award AST-0449558 to Vicky Kalogera at
Northwestern University. We thank the anonymous referee for comments
that helped to improve the presentation of the paper.

\end{document}